\documentclass{aa}  

\usepackage{graphicx}
\usepackage{xcolor}
\usepackage[version=4]{mhchem}
\usepackage{pdflscape}
\usepackage{txfonts}
\usepackage{hyperref}
\usepackage{cleveref}
\begin{document}

   \title{Complex organic molecules and cosmic ray ionisation rate \\ towards the massive protostar Cepheus A HW2}
   \titlerunning{COMs and CRIR towards Cep A}

   \author{Emma W. Nielsen
          \inst{1}\thanks{emma.w.nielsen@nbi.ku.dk}
          \and
          Anna Punanova\inst{2}
          \and
          Eva Wirström\inst{3}
          \and 
          Brandt Gaches\inst{4}
          \and \\
          A. O. Henrik Olofsson\inst{2}
          \and 
          Paola Caselli\inst{5}
          \and
          Prasanta Gorai\inst{6}
          \and 
          Jonathan C. Tan\inst{3,7}
          }

   \institute{Niels Bohr Institute, University of Copenhagen, Jagtvej 155 A, 2200 Copenhagen N, Denmark 
         \and
         Department of Physics and Astronomy, Chalmers University of Technology, Onsala Space Observatory, Observatoriev\"agen 90, R\aa\"o, 439 92, Onsala, Sweden
         \and
         Department of Physics and Astronomy, Chalmers University of Technology, 412 96, Gothenburg, Sweden
         \and
         Faculty of Physics, University of Duisburg-Essen, Lotharstraße 1, 47057 Duisburg, Germany    
         \and
         Max-Planck-Institut für Extraterrestrische Physik, Giessenbachstrasse 1, 85748 Garching, Germany
         \and
         Universität Heidelberg, Zentrum für Astronomie, Institut für Theoretische Astrophysik, Albert-Ueberle-Straße 2, 69120, Heidelberg, Germany
         \and
         Department of Astronomy, University of Virginia, Charlottesville, VA 22904-4235, USA
             }
 
  \abstract
  {Cosmic rays (CRs) are important drivers for molecular chemistry in star-forming regions, and laboratory experiments have shown that CRs can stimulate the release of complex organic molecules (COMs) such as methanol. Observationally, this has primarily been tested in cold, low-mass cores, so studying how CRs affect COM formation in a high-mass star-forming environment is of great interest. We performed a high-sensitivity wide-band spectral line survey with the Onsala 20 m telescope towards the high-mass protostar Cepheus A HW2, which is known to host an ionised jet. Consistent with previous studies, two primary velocity components ($-11$ km~s$^{-1}$ and $-5$ km~s$^{-1}$) were identified. Column densities and relative abundances of the detected ions and COMs were estimated from rotational diagrams, single transitions and RADEX grid searches (\ce{CH3OH}: $1.6\times10^{-9}$, \ce{CH3CN}: $5.9\times10^{-11}$, \ce{t-HCOOH}: $7.9\times10^{-11}$, \ce{H2CCO}: $1.7\times10^{-11}$, \ce{CH3CHO}: $1.9\times10^{-11}$, \ce{CH3OCHO}: $7.6\times10^{-10}$ at $-11$ km~s$^{-1}$). Deuterium fractions were also estimated (in range $0.002-0.3$ at $-11$~km~s$^{-1}$), and the volume density of molecular hydrogen ($2.6\times10^5$ cm$^{-3}$ at $-11$ km~s$^{-1}$) was constrained from the RADEX grid searches. Electron fractions and CR ionisation rates (CRIR, $6.8\times10^{-17}$ s$^{-1}$ at $-11$ km~s$^{-1}$, $\leq9.2\times10^{-19}$  s$^{-1}$ at $-5$ km~s$^{-1}$) were estimated through analytic chemistry using different ions as probes. The gas-grain chemical code {\sc nautilus} reproduced the observed abundances of \ce{CH3OH}, \ce{CH3CN}, \ce{HCO+}, \ce{N2H+} at the observed density, temperature and CRIR within the uncertainty of the model. The results indicate that the CR ionisation rate of the kinematic component associated with most of the COMs' emission in the region is locally enhanced. }


   \maketitle
%

\section{Introduction}
Energetic, charged particles, called cosmic rays (CRs), are important drivers for molecular chemistry in star-forming regions where external UV radiation is severely attenuated. Much of the chemistry is initiated by the ionisation of molecular hydrogen ($\ce{H2}+\ce{CR}\to\ce{H2+}+e^-$), which in turn produces \ce{H3+} that quickly reacts with abundant neutral species such as \ce{CO}, \ce{N2} and \ce{O} \citep[e.g.,][]{Herbst_1973,Viti2013,Gaches_2026}. As a result, the chemical complexity of the region is greatly enhanced. 

Complex organic molecules (COMs) can often be formed through ice-phase processes, that is, they are built up on grains and then desorbed into warmer gas \citep{Watanabe_2002,Garrod_2007}. Laboratory experiments have shown that CRs stimulate the release (sputtering) of COMs as simple as methanol \citep{Dartois2020} and as complex as glycine \citep{Esmaili2018}. 
Using simulations, \cite{Reboussin2014} and \cite{Shingledecker2018} further found that the gas phase abundance of COMs is increased at higher CR ionisation rates (CRIRs). Also using simulations, \cite{Caselli_2008}, \cite{Kong_2015} and \citet{Shingledecker2016} found that increasing the CRIR decreases the deuterium fractions of \ce{H2D+}/\ce{H3+}, \ce{N2D+}/\ce{N2H+} and \ce{DCO+}/\ce{HCO+}, respectively. Observationally, these predictions have primarily been tested in low-mass cores \citep[specifically L1174, TMC-1, L1544, L183;][]{Shingledecker2016,Shingledecker2018,Redaelli_2024}, where CRs are the primary source of energy, so the CR chemistry can be isolated. However, the CR flux is typically not enhanced in such regions, so studying how CRs affect the formation of COMs in high-mass star-forming regions (HMSFRs) that have potentially higher CRIRs is of great interest. 

To study how a possibly high CRIR affects the chemical complexity in star-forming regions, we examined Cepheus A HW2 (Cep A HW2), a high-mass protostar \citep[$12-24 M_\odot$, $10^4 L_\odot$; SOMA project, ][]{DeBuizer2017,Fedriani2023,Telkamp2025} resided in a nearby \citep[700 pc; ][]{Moscadelli_2009} HMSFR. Cep A HW2 is known to host an ionised jet \citep{Torrelles1993,Brogan2007,DeBuizer2017} and an ultra compact HII region \citep[UCHII][]{Hughes1984,Goetz1998}, suggesting that its CRIR could be enhanced; and COMs like methanol and methyl cyanide have already been detected towards the protostar \citep{Brogan2007}. The chemical complexity and CRIR in the region is, however, not well constrained. Thus, we performed an unbiased inventory of the chemical composition of Cep A HW2 and estimated the CRIR towards it using different ions as probes. We compared our results with previous works to examine how the CRIR affects the abundance of COMs as well as deuteration.

In Section~\ref{sec:observations_and_data_reduction}, we describe the observations and list the detected molecules and isotopologues. In Section~\ref{sec:results}, we present the kinematic components identified towards Cep A HW2 as well as the measured column densities, abundances, deuterium fractions, electron fractions and CRIRs. We discuss our results in Section~\ref{sec:discussion}, and we present our conclusions in Section~\ref{sec:conclusion}.
















\section{Observations and data reduction} \label{sec:observations_and_data_reduction}
Single-dish observations were carried out towards Cep A HW2 (J2000 $\alpha$: 22\textsuperscript{h}56\textsuperscript{m}17.97\textsuperscript{s}, $\delta$: $+62^\circ01'49.3''$) with the Onsala Space Observatory (OSO) 20~m telescope and OSA spectrometer during 2024 January-November and 2025 January-June (OSO project O2023b-06). The observations were carried out in position-switching mode with off-positions shifted $11'$ and $6'$ in azimuth for the 3-4~mm and 7~mm bands, respectively. Combining 22 setups, the observations covered the 3-4 mm (70-105, 108-113 GHz) and 7 mm (17-41 GHz) bands, both with a spectral resolution of 76~kHz, corresponding to velocity resolutions of 0.20-0.33~km~s$^{-1}$ and 0.63-1.3~km~s$^{-1}$, respectively. The system temperatures for the 3-4 mm and 7 mm bands were 220-470~K and 60-250~K, and the sensitivity was 5-15~mK [$T_{\rm mb}$]. Main beam efficiencies of 0.33-0.64 and 0.55-0.65 for the 3-4 mm and 7 mm bands, respectively, were applied according to the model that depends on observed frequency and elevation. The model obtained for 86~GHz was extrapolated over all bands using the Ruze relation. Planet observations indicate that the model should be accurate within 10\% at most elevations and frequencies. The half-power beam width (HPBW) was 33-$53''$ in the 3-4 mm band and  155-$157''$ for the \ce{NH3} transitions in the 7 mm band. The pointing and focus of the telescope were adjusted every 2 and 4 hours for the 3-4 mm and 7 mm bands, respectively. Table~\ref{tab:detected_molecules} lists the molecules we detected towards Cep A HW2 with a signal-to-noise (S/N) threshold of 3. In the remainder of this work, we analyse only COMs, ions and isotopologues of CO, \ce{NH3}, HCN and HNC. 

To reveal the source size and apply a correct beam-filling factor, additional observations were carried out on 22--23 October 2025 with the OSO 20~m telescope and OSA spectrometer over the 108.63--112.57~GHz band that provides the smallest beam size of $\simeq$33--34$^{\prime\prime}$. 21 pointing observations were done along the 170$^{\prime\prime}$-long strips with a step of 17$^{\prime\prime}$ (half-beam spacing), centred at Cep~A HW2, along RA and Dec axes, with a sensitivity of 150-170~mK (in $T_{\rm mb}$ scale). The band covers four bright spectral lines that trace gas of different density: \ce{C^18O} $(1-0)$ and \ce{C^17O} $(1-0)$ trace extended molecular gas, \ce{SO} $(2_3-1_2)$ traces dense gas associated with methanol, \ce{HC3N} $(12-11)$ traces the densest molecular gas.    

\begin{table*}[]
    \caption{Molecules detected towards Cep A HW2.} 
    \label{tab:detected_molecules}
    \centering
    \begin{tabular}{l|l}
    \hline \hline 
COMs & \ce{CH3CN}, \ce{CH3OH}, \ce{CH3CHO}, \ce{CH3OCHO}, \ce{H2CCO}, \ce{t-HCOOH}, \ce{CH3CCH} \\ 
Ions & \ce{N2H+}, \ce{N^15NH+}, \ce{HCO+}, \ce{DCO+}, \ce{H^13CO+}, \ce{HC^18O+}, \ce{HCS+}, \ce{CF+} \\ 
Simple neutrals & \ce{NH3}, \ce{NH2D}, \ce{HCN}, \ce{DCN}, \ce{H^13CN}, \ce{HC^15N}, \ce{HNC}, \ce{DNC}, \ce{H^15NC}, \ce{HN^13C}, \ce{^13C^18O}, \ce{C^18O}, \ce{^13CO}, \ce{C^17O}, \\
& \ce{H2O}, \ce{HDO}, \ce{^13CN}, \ce{HNCO}, \ce{HCO}, \ce{H2CO}, \ce{H2^13CO}, \ce{CCH}, \ce{SiO} \\ 
Cyanopolyynes & \ce{HC3N}, \ce{H^13CCCN}, \ce{HC^13CCN}, \ce{HCC^13CN}, \ce{HC5N} \\
S-bearing & \ce{OCS}, \ce{H2CS}, \ce{CS}, \ce{^13CS}, \ce{C^34S}, \ce{C^33S}, \ce{CCS}, \ce{SO}, \ce{^34SO}, \ce{SO2} \\ \hline
    \end{tabular}
\end{table*}

\section{Analysis and results} \label{sec:results}
\subsection{Line profiles}
\label{subsec:line_fitting}
The spectra were calibrated and averaged with \textsc{gildas} software.\footnote{\textsc{gildas}: \url{https://www.iram.fr/IRAMFR/GILDAS/}} Hyperfine structures (HFSs) were fitted using the python package \textsc{pyspeckit} \citep{Ginsburg_2011,Ginsburg_2022}. Assuming the excitation temperature is the same for all hyperfine components, \textsc{pyspeckit} calculates the HFS from the relative velocities and intensities of the hyperfine components. The HFS fits constrained the local standard of rest velocity $V_{\rm LSR}$, velocity dispersion $\sigma$ or ${\rm FWHM}=2\sqrt{2\ln{2}} \sigma$, peak temperature $T_{\rm peak}$, excitation temperature $T_{\rm ex}$ and optical depth $\tau$. Gaussian profiles were fitted to the lines with no HFS, constraining only $V_{\rm LSR}$, $\sigma$ and $T_{\rm peak}$. If two distinct velocity components were visually identified, a sum of two profiles with independent parameters was fitted instead. Nearby emission lines were masked out, and a third degree polynomial was fitted to subtract the baseline. The studied line profiles are shown in Figs.~\ref{fig:methanol_peaks}-\ref{fig:hyperfine_fits}. While detected, the methanol lines at 96.741 and 96.744 GHz were not studied because they are severely blended with another component with $V_{\rm LSR}\simeq-14$ km~s$^{-1}$.

All of the detected COMs, ions and simple neutrals show emission at $V_{\rm LSR}=(-10.56\pm0.06)$ km~s$^{-1}$, while methanol and methyl cyanide also show emission at $(-4.73\pm0.09)$ km~s$^{-1}$ (mean velocities were calculated from the detected methanol lines; see Tables~\ref{tab:detected_COM_lines} and \ref{tab:detected_lines} for the fitted velocities and line widths). Similar velocity components were reported by, e.g., \cite{Brogan2007}. For methanol, the emission is stronger at $-11$ km~s$^{-1}$ than at $-5$ km~s$^{-1}$ except at 80.99, 89.51 and 104.30 GHz, where $E_u$ is high (see Table \ref{tab:detected_COM_lines} for the energies of the methanol lines).

Some ions and simple neutrals exhibit an additional splitting of the component at $-11$ km~s$^{-1}$ (marked in Fig. \ref{fig:hyperfine_fits} with a green plus). These secondary components have $V_{\rm LSR}=(-11.4\pm0.2)$ km~s$^{-1}$ and $(-9.79\pm0.09)$ km~s$^{-1}$, that is, they are shifted by $\sim0.8$~km~s$^{-1}$ (see Table~\ref{tab:two-components-10} for $V_{\rm LSR}$ and $\sigma$). Except for \ce{^13CO}, the component at $-10$ km~s$^{-1}$ is the strongest. The average velocity dispersions of the $-11.4$ and $-9.8$ km~s$^{-1}$ components agree within 2 sigma. 

\subsubsection{Source size}\label{sec:source_size}
\begin{figure}
    \centering
    \includegraphics[width=\linewidth]{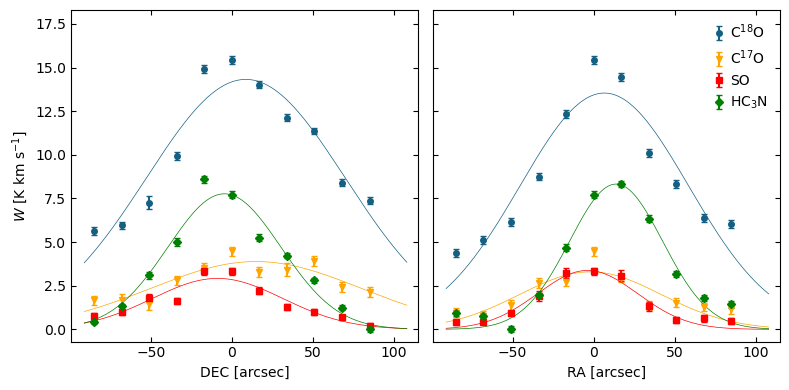}
    \caption{Gaussian fits to the integrated intensities observed at different positions along RA and Dec axes. The (0,0) position corresponds to Cep~A HW2. }
    \label{fig:source_sizes}
\end{figure}

The source size of Cep A HW2 was estimated from cross-maps of \ce{C^18O} $(1-0)$, \ce{C^17O} $(1-0)$, \ce{SO} $(2_3-1_2)$ and \ce{HC3N} $(12-11)$ (see Sect. \ref{sec:observations_and_data_reduction}). Gaussian profiles were fitted to the integrated intensities measured at each position, see Table \ref{tab:source_sizes} and Fig.~\ref{fig:source_sizes}. Assuming that \ce{H2}, COMs and ions emit from the same volume of dense gas, we estimate the source size to be $\sim80''$ or 0.27 pc (an average of \ce{SO}, known to trace \ce{CH3OH}, and of \ce{HC3N}, known to trace dense gas). The source size of HCN, HNC and their isotopologues ($\sim107''$ or 0.36 pc) was taken as an average of \ce{CO} and \ce{SO}. 

\begin{table}
\caption{Source sizes (the Gaussian fits to the integrated intensities). }
    \label{tab:source_sizes}
    \centering
    \begin{tabular}{lrr}
    \hline \hline 
       Line & FWHM (Dec) [$''$] & FWHM (RA) [$''$] \\ \hline
       \ce{C^18O}(1--0) & $144\pm2$ & $123\pm1$ \\
       \ce{C^17O}(1--0) & $153\pm10$ & $103\pm6$ \\
       \ce{SO}(2$_3$--1$_2$) & $95\pm4$ & $73\pm 3$ \\
       \ce{HC3N}(12--11) & $82\pm1$ & $68\pm1$ \\ \hline
    \end{tabular}
\end{table}

\subsection{Column densities}
\subsubsection{Rotational diagrams}
\label{subsec:rotational_diagram}
Rotational diagrams were used to estimate the column densities of methanol (\ce{CH3OH}) E/A, methyl cyanide (\ce{CH3CN}), acetaldehyde (\ce{CH3CHO}) E/A, ketene (\ce{H2CCO}) and trans-formic acid (\ce{t-HCOOH}), each of which had several detected transitions. Assuming thermalised excitation and that the emission is optically thin, the column density of the upper state $N_u$ was calculated as
\begin{align}
    N_u=\frac{8\pi k\nu^2 W}{hc^3A}\frac{\theta_s^2+\theta_b^2}{\theta_s^2},
\end{align}
where $k$ is the Boltzmann constant, $\nu$ is the frequency, $h$ is the Planck constant, $c$ is the speed of light, $A$ is the Einstein coefficient, $W$ is the velocity-integrated intensity of the line, $\theta_s$ is the source FWHM (for the COMs, we adopt $\sim80^{\prime\prime}$ source size, see Sect.~\ref{sec:source_size}), and $\theta_b$ is the HPBW \citep[e.g.,][the parameters are given in Table~\ref{tab:detected_COM_lines}]{Goldsmith_1999}. $\theta_s^2/(\theta_s^2+\theta_b^2)$ is the beam filling factor, assuming the source is Gaussian \citep{Huttemeister_1993}. 

Assuming LTE conditions, the total column density $N$ is related to $N_u$ by
\begin{align}
    \ln\left(\frac{N_u}{g_u}\right) = \ln\left(\frac{N}{Q_\text{rot}}\right) - \frac{E_u}{kT_\text{rot}},
\end{align}
where $g_u$ is the upper state degeneracy, $Q_\text{rot}$ is the rotational partition function, $E_u$ is the upper state energy (given in Table~\ref{tab:detected_COM_lines}), and $T_\text{rot}$ is the rotational temperature \citep[e.g.,][]{Goldsmith_1999}. To estimate $N$, a linear function was fitted to $\ln(N_u/g_u)$ versus $E_u/k$. $Q_\text{rot}(T_\text{rot})$ was fitted through a third or fifth degree polynomial to the values of $Q_{\rm rot}(T)$ listed at CDMS/JPL \citep{Pickett1998,Muller2001,Endres2016}. The applied $T_{\rm rot}$ and $Q_{\rm rot}$ are listed in Table~\ref{tab:column_density_abundance}. Known methanol maser lines \citep[at 24928.715, 24959.084, 36169.29, 84521.206 and 95169.516 MHz;][]{M_ller_2004} were excluded because their excitations are not only collisional. 

\begin{figure}
    \centering
    \includegraphics[width=\linewidth]{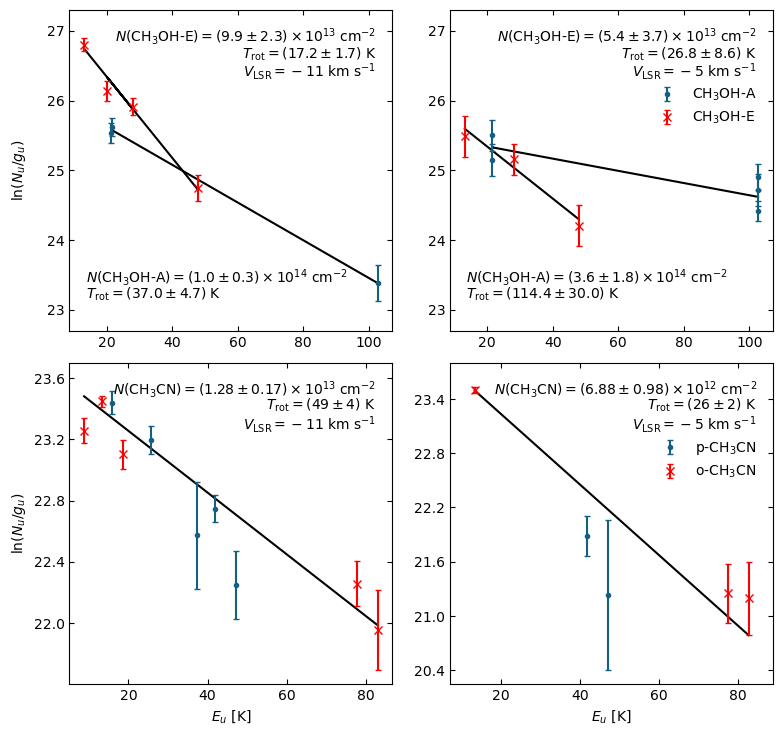}
    \caption{Rotational diagrams at $V_{\rm LSR}=-11$ km~s$^{-1}$ (left) and $-5$ km~s$^{-1}$ (right). \textit{Top:} Methanol: \ce{CH3OH}-E (blue dots) and \ce{CH3OH}-A (red crosses). \textit{Bottom:} Methyl cyanide: p-\ce{CH3CN} (blue dots) and o-\ce{CH3CN} (red crosses).}
    \label{fig:rtd_methanol_ch3cn}
\end{figure}

Only low-energy ($E_u<50$ K) methanol E lines were included in the rotational diagrams in Fig.~\ref{fig:rtd_methanol_ch3cn} because the low- and high-energy lines have different excitation conditions (see Sect.~\ref{subsec:radex_grid_search}). We did not have enough low-energy methanol A lines to distinguish between low- and high-energy excitation conditions, so the presented rotational temperatures and column densities of methanol A may be overestimated\footnote{Including high-energy methanol E lines in the rotational diagram for methanol E increased the rotational temperature and column density of methanol E. }. Because of the non-thermal nature of methanol excitation and to probe volume density and kinetic temperature, we also modelled $N(\ce{CH3OH})$ with RADEX and used that result for further analyses (see Sect.~\ref{subsec:radex_grid_search}).

\begin{figure}
    \includegraphics[width=\linewidth]{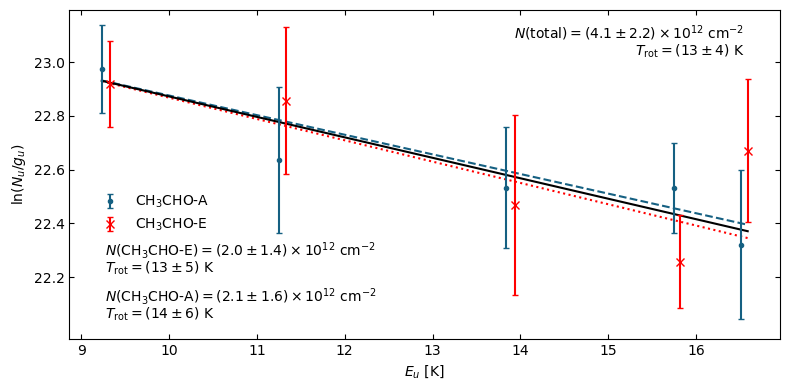}
    \includegraphics[width=\linewidth]{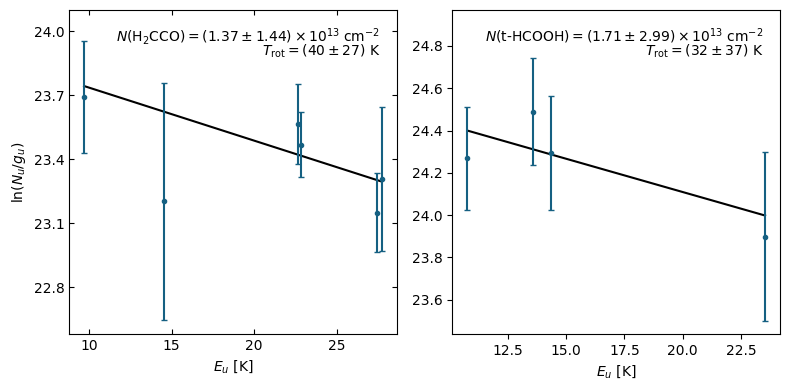}
    \caption{Rotational diagrams at $V_{\rm LSR}=-11$ km~s$^{-1}$. \textit{Top:} Acetaldehyde: \ce{CH3CHO}-A (blue dots) and \ce{CH3CHO}-E (red crosses). Individual fits (blue dashed and red dotted lines for \ce{CH3CHO}-A and \ce{CH3CHO}-E, respectively) and a combined fit (black line) are shown. \textit{Bottom:} Ketene (\ce{H2CCO}) (left) and trans-formic acid (t-HCOOH) (right). }
    \label{fig:rtd_acetaldehyde_ketene_thcooh}
\end{figure}

The column densities of methanol and methyl cyanide were calculated separately for each velocity component (see Fig.~\ref{fig:rtd_methanol_ch3cn}); the column density of acetaldehyde -- for molecular forms E and A. Based on the individual fits, acetaldehyde has an E/A ratio of $\sim1$, so the detected acetaldehyde E/A lines were combined to better constrain the total column density of acetaldehyde (see the upper panel of Fig.~\ref{fig:rtd_acetaldehyde_ketene_thcooh}). The rotational diagrams of ketene and trans-formic acid are shown in the lower panels of Fig.~\ref{fig:rtd_acetaldehyde_ketene_thcooh}. The column densities are summarised in Table \ref{tab:column_density_abundance}. Methanol is the most abundant COM towards Cep A HW2 with $N=(3.52\pm0.08)\times10^{14}$~cm$^{-2}$. Methyl cyanide ($N\sim1.3\times10^{13}$~cm$^{-2}$) and ketene ($N\sim1.4\times10^{13}$~cm$^{-2}$) are $\sim30$ times less abundant. Trans-formic acid ($N\sim1.7\times10^{13}$~cm$^{-2}$ from rotational diagram, see also Sect. \ref{subsec:thin_thick_estimates}) is $\sim20$ times less abundant, and acetaldehyde ($N\sim4.1\times10^{12}$~cm$^{-2}$) is $\sim86$ times less abundant.

\subsubsection{Optically thin/thick transitions}
\label{subsec:thin_thick_estimates}
When only one transition was detected for a given molecule, the column density was calculated as in \cite{Caselli_2002}. For optically thick transitions with HFSs,
\begin{align}
    N=\frac{8\pi^{3/2}\sigma}{g_u \lambda^3 A} \frac{\tau}{1-e^
    {-h\nu/kT_{\rm ex}}}\frac{Q_{\rm rot}}{e^{-E_l/kT_{\rm ex}}}\frac{\theta_s^2+\theta_b^2}{\theta_s^2},
\end{align}
where $\tau$ is the optical depth, and $E_l$ is the lower state energy. For linear molecules, we calculated
\begin{align}
    Q_{\rm rot}=\sum_{J=0}^{99}(2J+1)e^{-E_J/kT_{\rm ex}},
\end{align}
where $E_J=J(J+1)hB$, $J$ is the rotational quantum number, and $B$ is the rotational constant \citep{Caselli_2002}. For non-linear molecules, $Q_{\rm rot}$ was retrieved from CDMS \citep{Muller2001,Endres2016}. 

If a transition had no HFS or $\tau\leq0.1$, the transition was assumed to be optically thin, and the column density was calculated as
\begin{align} \label{eq:column_density_thin}
    N=\frac{8\pi W}{\lambda^3 A g_u}\frac{1}{J_\nu(T_{\rm ex})-J_\nu(T_{\rm CMB})}\frac{1}{1-e^
    {-h\nu/kT_{\rm ex}}}\frac{Q_{\rm rot}}{e^{-E_l/kT_{\rm ex}}}\frac{\theta_s^2+\theta_b^2}{\theta_s^2},
\end{align}
where $J_\nu(T_{\rm ex})$ and $J_\nu(T_{\rm CMB})$ are the Rayleigh-Jeans equivalent excitation and background temperatures. Here, we adopted $T_{\rm ex}=4.59$ K from \ce{H^13CO+} (Table \ref{tab:column_density_abundance}) for ions and simple neutrals and, assuming the emission of COMs is thermalised, $T_{\rm ex}=33$~K from methanol (Table \ref{tab:radex_results}) for COMs. The secondary components at $-11.4$~km~s$^{-1}$ and $-9.8$ km~s$^{-1}$ had poorly constrained $T_{\rm ex}$ and $\tau$, so we only estimated the total column density at $-11$ km~s$^{-1}$. The column density of \ce{NH3} was obtained directly from an HFS fit of the 1--1, 2--2 and 3--3 transitions assuming an ortho/para ratio of 1/1 \citep{Harju_2017}. The fit gave $T_{\rm rot}=(21.9\pm1.3)$ K, which is smaller than $T_{\rm rot}$ inferred from the rotational diagrams and $T_{\rm kin}$ from RADEX grid search, but similar to $T_{\rm kin}=19.5-28.6$~K inferred from the ratio of the integrated intensities of HCN/HNC \citep[$T_{kin}\simeq10\times W_{\ce{HCN}}/W_{\ce{HNC}}$, e.g.,][]{Hacar_2020,Pazukhin_2022,Pazukhin2023}.

$3\sigma$ upper limits on the column densities of some non-detected species (\ce{CH3OCH3} and \ce{NH2CHO} at $-11$ km~s$^{-1}$; \ce{N2H+} and \ce{HC^18O+} at $-5$ km~s$^{-1}$) were calculated from Eq.~\ref{eq:column_density_thin} by substituting 3 rms $\times$ 2 km~s$^{-1}$ for $W$, where the rms was calculated in the region where the line would have been, and 2 km~s$^{-1}$ is the assumed length of the integration interval. For \ce{N2H+} and \ce{HC^18O+}, we instead adopted the line widths at $-11$ km~s$^{-1}$. 

The results are shown in Table~\ref{tab:detected_lines}. To infer the column densities of the main isotopologues (see Table~\ref{tab:column_density_abundance}), we used the isotope ratios $\ce{^12C}/\ce{^13C}=70$, $\ce{^16O}/\ce{^18O}=523$ and $\ce{^14N}/\ce{^15N}=446$ calculated cf. \cite{Wilson_1994} given the Galactocentric distance of Cep A is 8.26~kpc \citep{Bobylev_2010}. For molecules where several isotopologues correspond to a range of estimates, we adopted the largest ones for further calculations because we assume that a lower column density may indicate optically thick emission. 
$N(\ce{N2H+})$ inferred from the \ce{N^15NH+} (1--0) transition is about 4 times larger than the direct measurement from the \ce{N2H+} (1--0) transition. However, as shown in Fig.~\ref{fig:hyperfine_fits}, the \ce{N^15NH+} (1--0) transition is poorly modelled by its hyperfine fit. The column density of trans-formic acid was estimated both using a rotational diagram and by making an optically thin estimate of its brightest line, and the results agree well within $1\sigma$. 

\subsubsection{RADEX grid search}
\label{subsec:radex_grid_search}
The column density of methanol and the volume density of molecular hydrogen were estimated from a RADEX grid search. RADEX \citep{van_der_Tak_2007} is a one-dimensional non-LTE radiative transfer code that calculates the intensities of selected lines. We used the python implementation \textsc{pythonradex} \citep{Cataldi_pythonradex} to calculate the intensities and excitation temperatures of the detected methanol lines over a $30\times30\times30$ grid of logarithmically spaced \ce{H2} volume densities, $n(\ce{H2})$, and methanol column densities, $N(\ce{CH3OH})$, and linearly spaced kinetic temperatures, $T_{\rm kin}$. The initial grids spanned $\log[n(\ce{H2})]=3.5-12$, $\log[N(\ce{CH3OH})]=11.5-16.5$ and $T=10-250$ K, and the ranges were adjusted iteratively to resolve the uncertainties on the fitted parameters. The final ranges are presented below.  

The thermalised o/p ratio of \ce{H2} in molecular clouds is very small \citep[$10^{-4}-10^{-2}$ cm$^{-3}$, ][]{Furuya2015}, so the collisional effect of \ce{o-H2} is negligible. We therefore assumed the o/p ratio of \ce{H2} to be 0, and that \ce{p-H2} was the only collisional partner; the collisional data was retrieved from \cite{Dagdigian2024}. We further assumed the source to be a uniform sphere, and that the temperature of the dust continuum radiation field was 0 K. We assumed the line profiles to be Gaussian with line widths equal to the average measured line width of the inputted lines ($\sigma$ in range $1.4-1.7$ km~s$^{-1}$).

\textsc{pythonradex} calculates the line peak intensity $S$ in W m$^{-2}$ Hz$^{-1}$, which was converted to a peak main beam temperature $T_{\rm mb}$ using
\begin{align}
    T_{\rm mb}=\frac{\lambda^2}{2k\Omega}S,
\end{align}
where $\lambda$ is the wavelength, and $\Omega=\pi\theta_s^2/(4\ln2)$ is the solid angle of the source. The integrated intensity was then calculated as $W=\sqrt{2\pi}\sigma T_{\rm mb}$. \textsc{pythonradex} does not take the beam size into account, so the predicted peak main beam temperatures were rescaled by the beam filling factor in order to compare the predicted and measured integrated intensities:
\begin{align}
    T_{\rm mb, measured} = \frac{\theta_s^2}{\theta_s^2+\theta_b^2} T_{\rm mb, radex}.
\end{align}
The grid of predicted integrated intensities was converted to a three-dimensional grid of $\chi^2$, and the best-fit parameters were selected to minimise the $\chi^2$. The 1, 2 and $3\sigma$ regions of $n(\ce{H2})$, $N(\ce{CH3OH})$ and $T_{\rm kin}$ satisfy $\chi^2-\chi^2_{\rm min}\leq3.51$, 7.82 and 13.93, respectively, where $\chi^2_{\rm min}$ is the minimum $\chi^2$. The $1\sigma$ confidence region of each parameter taken separately satisfies $\chi^2-\chi^2_{\rm min}\leq1$, and the uncertainties of the fitted parameters were found by projecting this region onto each of the parameter axes. 

We initially used all 10 methanol E and A lines at $-11$ km~s$^{-1}$ and all 12 lines at $-5$ km~s$^{-1}$ for the fits. However, with the model implemented in RADEX, it was not possible to fit all of the lines with just one $T_{\rm kin}$, $n(\ce{H2})$ and $N(\ce{CH3OH})$ for each velocity component (see Fig.~\ref{fig:grid_search_methanol_all_lines}). For the component at $-11$ km~s$^{-1}$, we then fitted just the low-energy lines ($E_u<50$ K), but the low-energy A methanol lines seemed to behave differently from the other low-energy lines. Excluding these lines, the fit converged. We tried including the line at 36 GHz, but this line also seemed to behave differently. To mimic the background radiation from Cep A HW2, we tried adding an external radiation field from a blackbody with temperatures $T_{\rm bkg}=5,10,20$ K. This worsened the fit significantly, so we proceeded with an external radiation field originating only from the CMB. For the component at $-5$ km~s$^{-1}$, we fitted the low- and high-energy lines separately. The results of the final grid searches are summarised in Table \ref{tab:radex_results}. 

\begin{table*}
    \caption{Results of RADEX grid searches. }
    \label{tab:radex_results}
    \centering
\begin{tabular}{lllllll}
    \hline \hline
$V_{\rm LSR}$ [km~s$^{-1}$] & $E_u$ [K] & $n(\ce{H2})$ [cm$^{-3}$] & $T_{\rm kin}$ [K] & $N(\ce{CH3OH})$ [cm$^{-2}$] & $N(\ce{H2})$ [cm$^{-2}$] & $\chi^2$ \\
\hline
$-11$ & $<50$ & $(2.6_{-0.6}^{+0.8})\times10^5$ & $33\pm2$ & $(3.52\pm 0.08)\times10^{14}$ & $(2.2\pm0.6)\times10^{23}$ & 0.042 \\
$-5$ & $<50$ & $\sim10^7$ & $23_{-2}^{+5}$ & $(1.4\pm0.1)\times10^{14}$ & $\sim8\times10^{24}$ & 2.019 \\
$-5$ & $>50$ & $\sim10^8$ & $\gtrsim 130$ & $(5.4\pm0.5)\times10^{14}$ & $\sim 8\times10^{25}$ & 4.825 \\
\hline
\end{tabular}
\tablefoot{$N(\ce{CH3OH})$ is the total fitted column density of E and A methanol, assuming an E:A ratio of 1. The column density of \ce{H2} at $-5$ km~s$^{-1}$ was calculated assuming that the source size is the same as at $-11$ km~s$^{-1}$. }
\end{table*}

\begin{figure*}
    \centering
    \includegraphics[width=0.9\linewidth]{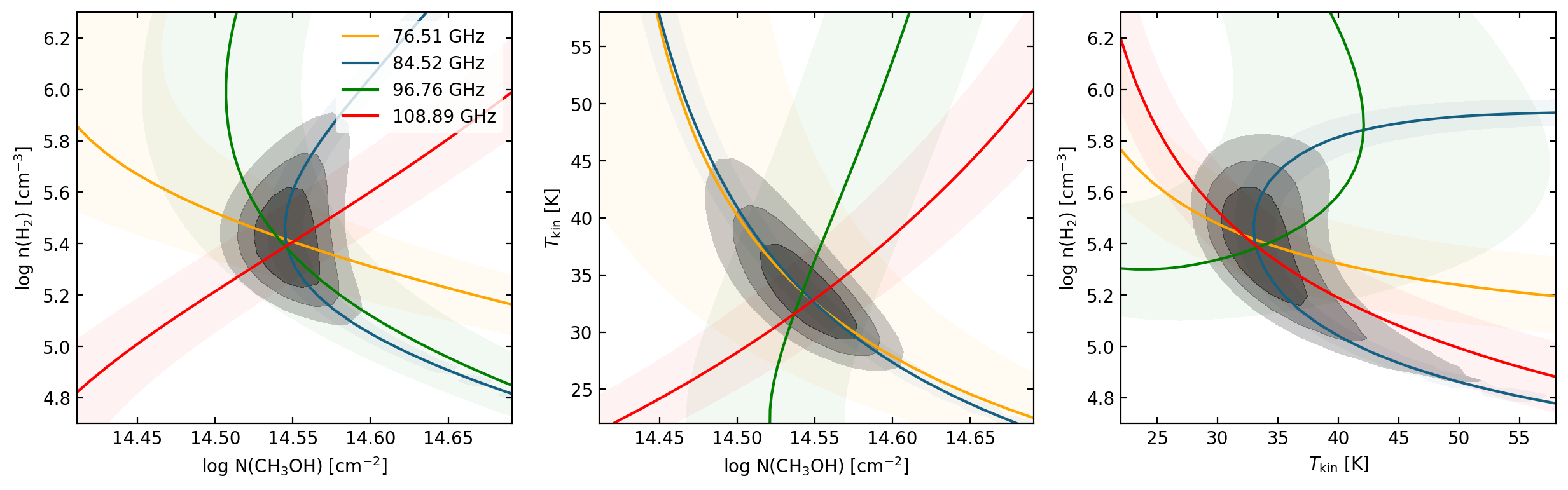}
    \caption{RADEX grid search using low-energy ($E_u<50$ K) methanol E lines at $-11$ km~s$^{-1}$. The coloured lines (and $1\sigma$ shaded regions) show what combinations of parameters are consistent with each of the measured integrated intensities. The grey regions show the 1, 2 and $3\sigma$ regions of the fitted parameters in the plane of the best fit $T_{\rm kin}$, $n(\ce{H2})$ and $N(\ce{CH3OH})$ (from left to right). The excitation temperatures for the lines at 76.51, 84.52, 96.74, 96.76 and 108.89 GHz are 17, $-7$, 17  and 10 K, respectively. }
    \label{fig:grid_search_methanol_11_E}
\end{figure*}

Figure~\ref{fig:grid_search_methanol_11_E} shows the result of a grid search using low-energy E methanol lines at $-11$ km~s$^{-1}$. The grid spanned $4.7\leq\log n(\ce{H2})\leq6.3$, $14.11\leq \log N(\ce{CH3OH-E})\leq14.39$ and $22\text{ K}\leq T_{\rm kin}\leq58$~K, and the $\chi^2$ is small. The best-fit parameters, which will be used for later analyses, are $n(\ce{H2})=(2.6_{-0.6}^{+0.8})\times10^5$~cm$^{-3}$, $N(\ce{CH3OH})=(3.52\pm 0.08)\times10^{14}$~cm$^{-2}$ and $T_{\rm kin}=(33\pm2)$~K. High-energy lines were excluded because they behaved differently: Fig.~\ref{fig:grid_search_methanol_all_lines} shows a grid search including all E/A methanol lines at $-11$ km~s$^{-1}$, but the line at 104.3~GHz with $E_u=158.6$~K cannot be modelled with parameters even close to the ones that well reproduce the low-energy lines. Thus, the high-energy lines were excited under different conditions, suggesting that they originate from a different volume. There were too few high-energy E/A methanol lines to make a separate constraint on $n(\ce{H2})$, $N(\ce{CH3OH})$ and $T_{\rm kin}$. The E methanol line at 96.74 GHz was excluded from the grid search in Fig.~\ref{fig:grid_search_methanol_11_E} because it is blended with neighbouring lines and maybe with the other velocity component too. The best-fit kinetic temperature differs from the rotational temperature of E methanol shown in Fig.~\ref{fig:rtd_methanol_ch3cn}, but this is expected as the excitations are not only collisional. For example, the line at 84.5 GHz is masing, so \textsc{pythonradex} predicts an excitation temperature of $-7$ K. 

\begin{figure*}
    \centering
    \includegraphics[width=0.9\linewidth]{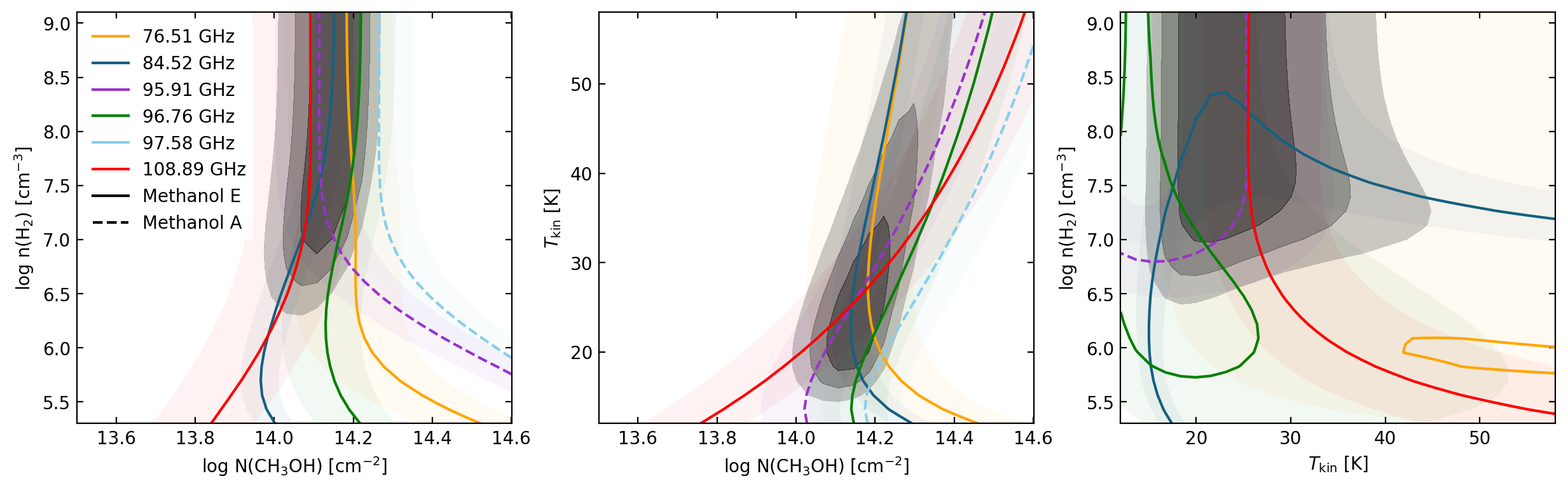}
    \caption{RADEX grid search using low-energy ($E_u<50$ K) methanol E/A lines at $-5$ km~s$^{-1}$. Coloured lines and shaded areas are as in Fig.~\ref{fig:grid_search_methanol_11_E}. The excitation temperatures for the lines at 76.51, 84.52, 95.91, 96.76, 97.58 and 108.89 GHz are 19, 31, 23, 20, 23 and 23 K, respectively. }
    \label{fig:grid_search_methanol_5_low_energy}
\end{figure*}

Figure~\ref{fig:grid_search_methanol_5_low_energy} shows the result of a grid search using a combination of low-energy ($E_u<50$ K) E and A methanol lines at $-5$ km~s$^{-1}$, assuming an E/A ratio of 1. The grid spanned $5.3\leq \log n(\ce{H2}) \leq9.1$, $13.2\leq \log N(\ce{CH3OH}) \leq14.3$ and $12\text{ K}\leq T_{\rm kin}\leq 58$ K, and the $\chi^2$ is close to the degrees of freedom. The emission is in LTE because all the integrated intensities are independent of $n(\ce{H2})$ above $\sim 10^7$~cm$^{-3}$. As a result, the observed emission can be explained by $n(\ce{H2})\sim 10^7$~cm$^{-3}$. The excitation temperatures of the low-energy lines are similar, and the kinetic temperature agrees with the rotational temperature for E methanol, see Fig.~\ref{fig:rtd_methanol_ch3cn} (the error on the latter is, however, $\sim30\%$). This is expected if the emission is thermal. 

\begin{figure*}
    \centering
    \includegraphics[width=0.9\linewidth]{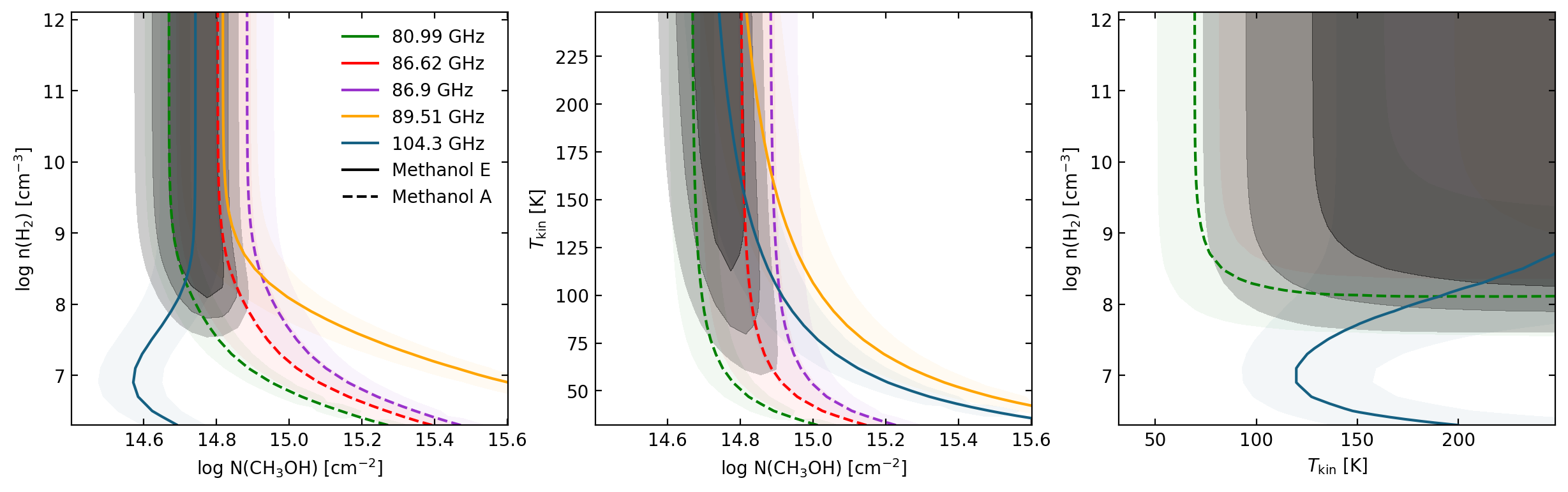}
    \caption{RADEX grid search using high-energy ($E_u>50$ K) methanol E/A lines at $-5$ km~s$^{-1}$ (excluding the line at 95.17 GHz). Coloured lines and shaded areas are as in Fig.~\ref{fig:grid_search_methanol_11_E}. The excitation temperatures for the lines at 80.99, 86.62, 86.9, 89.51, 104.3 GHz are 247, 248, 248, 247 and 247 K, respectively. }
    \label{fig:grid_search_methanol_5_high_energy}
\end{figure*}

Figure~\ref{fig:grid_search_methanol_5_high_energy} shows the result of a grid search using a combination of high-energy ($E_u>50$ K) E and A methanol lines at $-5$ km~s$^{-1}$, again assuming an E/A ratio of 1. The grid spanned $6.3\leq \log n(\ce{H2}) \leq12.1$, $14.1\leq \log N(\ce{CH3OH}) \leq15.3$ and $32\text{ K}\leq T_{\rm kin}\leq 248$ K (the maximum allowed $T_{\rm kin}$ in \textsc{pythonradex} is 250 K). The line at 95.17 GHz was excluded because it behaved differently, specifically, it seemed to be excited at lower $n(\ce{H2})$. The integrated intensities become independent of $n(\ce{H2})$ above a volume density of $\sim10^8$~cm$^{-3}$, again suggesting the emission is in LTE. The density and kinetic temperature were poorly constrained by the fit. 

The volume density of \ce{H2} was converted to a column density $N(\ce{H2})=n(\ce{H2})\;\times$ distance to Cep A $\times$ diameter of Cep A [rad], assuming the distance to Cep A is 700~pc \citep{Moscadelli_2009} and the size of Cep A is $80''$ (see Sect. \ref{sec:source_size}). The column density of \ce{H2} associated with the low-energy E methanol lines at $-11$ km~s$^{-1}$ is $N(\ce{H2})=(2.2\pm0.6)\times10^{23}$~cm$^{-2}$, consistent with $N(\ce{H2})>1.5\times10^{23}$~cm$^{-2}$ traced through silicates with Spitzer \citep{Sonnentrucker2008}. Further assuming that Cep A is a sphere primarily consisting of \ce{H2}, a crude estimate of its mass is $133 M_\odot$. This is similar to the core mass $M_c\sim120M_\odot$ calculated within $36.25''$ that was reported by \cite{Fedriani2023,Telkamp2025}. Our estimate is slightly higher, probably because the mass is calculated within a larger radius. The column density of \ce{H2} at $-5$ km~s$^{-1}$ was calculated assuming that the source size is the same as at $-11$ km~s$^{-1}$. The $-5$ km~s$^{-1}$ component could be more compact, but we do not know its size.

\subsection{Deuterium fraction}
The deuterium fraction $R_D$ is the column density ratio of deuterated to hydrogenated isotopologues. We calculated $R_D$ for all detected pairs of deuterated and hydrogenated species, namely \ce{DCO+}/\ce{HCO+}, DNC/HNC, DCN/HCN and \ce{NH2D}/\ce{NH3}. The column densities of \ce{HCO+}, HNC and HCN were inferred from the detected isotopologues (see Table~\ref{tab:column_density_abundance}); we used the largest estimates (inferred from \ce{HC^18O+}, \ce{HN^13C} and \ce{HC^15N}). The column density of \ce{NH2D} was inferred from a detected ortho-\ce{NH2D} transition assuming an ortho/para ratio of 3/1 \citep{Harju_2017}.

\begin{figure}
    \centering
    \includegraphics[width=0.9\linewidth]{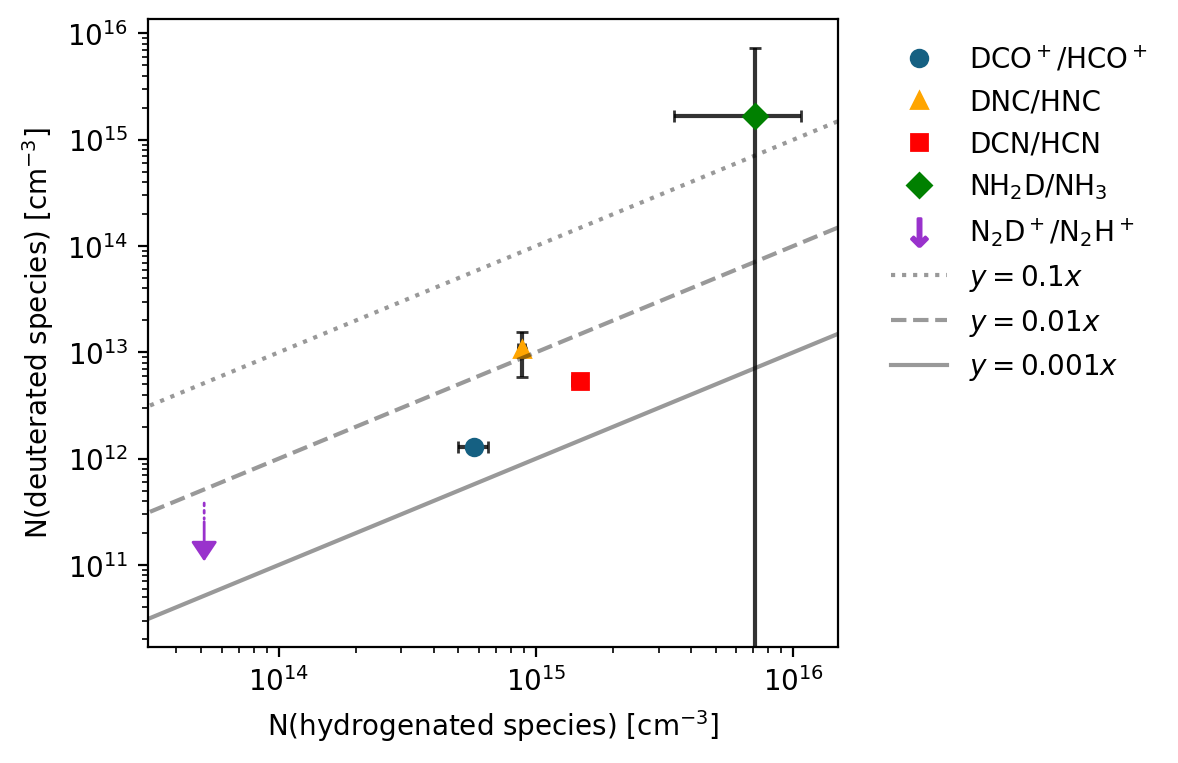}
    \caption{Column densities of deuterated against hydrogenated species. }
    \label{fig:deuterium_fraction}
\end{figure}

The deuterium fractions for the component at $-11$ km~s$^{-1}$ are shown in Table~\ref{tab:deuterium_fractions}, and Fig.~\ref{fig:deuterium_fraction} shows the column densities of the deuterated vs. hydrogenated molecules. \cite{Li_2017} reported $R_D^{\ce{DCO+}}=(7.9\pm1.6)\times10^{-4}$ and $R_D^{\ce{DCN}}=(2.4\pm0.5)\times10^{-3}$ from a single-dish observation towards Cep A with the IRAM 30~m telescope. Our measurements agree within 3.6 and 2.0~$\sigma$, respectively. The small discrepancy could be due to different beam sizes, assumptions about the beam filling factor and the isotopologue ratios. 

\begin{table}
    \caption{Deuterium fractions at $-11$ km~s$^{-1}$. }
    \label{tab:deuterium_fractions}
    \centering
\begin{tabular}{llr}
    \hline \hline
Deuterated & Hydrogenated & $R_D$ \\
\hline
\ce{DCO+} & \ce{HCO+} (\ce{HC^18O+}) & $(2.2 \pm 0.4)\times10^{-3}$ \\
\ce{DNC} & \ce{HNC} (\ce{HN^13C}) & $(1.2 \pm 0.6)\times10^{-2}$ \\
\ce{DCN} & \ce{HCN} (\ce{HC^15N}) & $(3.7 \pm 0.4)\times10^{-3}$ \\
\ce{NH2D} (\ce{oNH2D}) & \ce{NH3} & $(3.2 \pm 10.7)\times10^{-1}$ \\
\ce{N2D+} & \ce{N2H+} & $\leq 1.9\times10^{-3}$ \\
\hline
\end{tabular}
\tablefoot{Tracers are shown in parenthesis. }
\end{table}

Some transitions of deuterated versions of \ce{N2H+} (\ce{N2D+}), formaldehyde (\ce{HDCO}) and methanol (\ce{CH2DOH}) were covered by the frequency range of the spectrum but not detected. A $3\sigma$ upper limit of $R_D^{\ce{N2D+}}$ is listed in Table \ref{tab:deuterium_fractions}.

\subsection{Electron fraction and cosmic ray ionisation rate}
The electron fraction $x(e)$ and cosmic ray ionisation rate (CRIR) $\zeta$ were estimated through analytic chemistry using methods proposed by \cite{Caselli_1998} and \cite{Luo_2024}. Both methods consider the chemistry initiated when CRs ionise \ce{H2}, in turn producing ions such as \ce{H3+}, \ce{HCO+}, \ce{OH+} and \ce{H2O+}. \cite{Caselli_1998} also consider the formation of \ce{H2D+} and \ce{DCO+}, whereas \cite{Luo_2024} consider the formation of \ce{N2H+} and \ce{H3O+}. \cite{Caselli_1998} trace this chemistry from measurements of \ce{HCO+}, \ce{DCO+} and \ce{CO}, and \cite{Luo_2024} use \ce{N2H+} and \ce{HCO+} as tracers. We inferred $N(\ce{HCO+})$ from the measured \ce{HC^18O+} (1--0) transition and $N(\ce{CO})$ from the measured \ce{C^18O} (1--0) transition (see Table~\ref{tab:column_density_abundance}). 

The \cite{Caselli_1998} model is developed for dark cores and assumes that the reactions have reached a steady state. This model, which relies on measurements of deuterated species, further assumes that some CO is depleted, whereas the \cite{Luo_2024} model assumes that both depletion and deuteration are negligible. Cep A HW2 is not a dark core, but CO depletion and deuteration may not be negligible.

Following the method proposed by \cite{Caselli_1998}, we calculated
\begin{align}
    x(e)=\frac{3.3\times10^{-8}\text{ cm}^3/{\rm s}}{R_D^{\ce{DCO+}}} - \frac{2.6\times10^{-8}\text{ cm}^3/{\rm s}}{f_D}
\end{align}
and
\begin{align}
    \zeta = \left[2.7\times10^{-7}\text{ cm}^3/{\rm s}\;x(e) + \frac{9.0\times10^{-13}\text{ cm}^3/{\rm s}}{f_D} \right] x(e)n(\ce{H2})R_H,
\end{align}
where $R_H=N(\ce{HCO+})/N(\ce{CO})$, $f_D=f(\ce{^12CO})/f(\ce{^12CO})_{\rm fiducial}$ is the CO depletion factor, $f(\ce{X})$ is the relative abundance of species X with respect to \ce{H2}, and $f(\ce{^12CO})_{\rm fiducial}=2.7\times10^{-4}$ is the assumed CO abundance \citep{Lacy_1994}, if CO was not depleted \citep{Pineda_2024}. The numerical coefficients originally given in \cite{Caselli_1998} were updated to $3.3\times10^{-8}$, $2.6\times10^{-8}$, $2.7\times10^{-7}$ and $9.0\times10^{-13}\text{ cm}^3\;{\rm s}^{-1}$ by substituting reaction rates \citep{Caselli_1998,Bettens_1999,Woon_2009,Sipila_2018} assuming $T_{\rm kin}=20$~K (from \ce{NH3}) and HD abundance of $2.31\times10^{-5}$ \citep{Linsky_2006}; the reaction rates are available through the KIDA data base \citep{Wakelam_2012}.

Following the method proposed by \cite{Luo_2024}, we also calculated
\begin{align}
    x(e)=n(\ce{e-})/n(\ce{H2})
\end{align}
and
\begin{align}
    \zeta=n(\ce{H2})f(\ce{H3+})[&f(\ce{CO})k_{\ce{HCO+}}+f(\ce{N2})k_{\ce{N2H+}} \nonumber \\ &+f(\ce{O})k_{\ce{OH+,H2O+}}+f(\ce{e-})k_{\ce{H}}],
\end{align}
where $n(\ce{e-})$ is the volume density of electrons, and $k_{\rm X}$ is the rate of the reaction(s) that form(s) species X. $n(\ce{e-})$ and $n(\ce{H3+})$ were calculated using eq. (5)-(10) in \cite{Luo_2024} assuming $T_{\rm kin}=33$~K (from RADEX search), $f(\ce{O})=2.51\times10^{-4}$, $f(\ce{N2})=3.8\times10^{-5}$ \citep[from ][]{Luo_2024}. We performed the calculations using both the measured $f(\ce{CO})$ (where some of the CO is depleted) and $f(\ce{CO})=1\times10^{-4}$ (where no CO is depleted) taken from \cite{Luo_2024}.

In addition, a lower bound on the electron fraction was calculated from the detected ions as described in \cite{Caselli_2002}:
\begin{align}
    x(e)\geq\bigl[&N(\ce{HCO+}) + N(\ce{DCO+}) + N(\ce{N2H+}) + N(\ce{HC^18O+}) \nonumber \\ &+ N(\ce{H^13CO+})\bigr]/N(\ce{H2}).
\end{align}
The results are shown in Table \ref{tab:X(e)_and_CRIR}. The CRIR estimated using the method from \cite{Caselli_1998} is unrealistically high; it is of order $10^{-13}$~s$^{-1}$, whereas \cite{Caselli_1998} reported CRIRs in the range $10^{-18}-10^{-16}$~s$^{-1}$ for 24 dark cores. When testing the \cite{Caselli_1998} method against synthetic and real observations, \cite{Redaelli_2024} found that the CRIR was overestimated by $2-4$ orders of magnitude ($\zeta_{\rm estimate}\sim10^{-15}-10^{-13}$~s$^{-1}$, which is similar to our measurement towards Cep A HW2), despite updating the reaction rates in the original equations and separating ortho and para state reactions. 

For the component at $-11$~km~s$^{-1}$, our estimates using the method from \citet{Luo_2024} are $x(e)=(1.55\pm0.04)\times10^{-8}$ and $\zeta=(6.8\pm0.5)\times10^{-17}$~s$^{-1}$ (assuming $f_{\ce{CO}}=1\times10^{-4}$). 
An upper limit on the CRIR at $-5$ km~s$^{-1}$ was estimated from the $3\sigma$ upper limits on the column densities of \ce{N2H+} and \ce{HC^18O+}: $\zeta\leq9.2\times10^{-19}$~s$^{-1}$, which is not elevated. At a fixed $f_{\rm CO}$, the CRIR decreases slightly when the volume density of \ce{H2} increases, but it is relatively stable (see Table~\ref{tab:X(e)_and_CRIR}). 

\begin{table*}
    \caption{Electron fraction and CRIR estimates. }
    \label{tab:X(e)_and_CRIR}
    \centering
\begin{tabular}{lllll}
\hline \hline
Method & $n(\ce{H2})$ [cm$^{-3}$] & $f_{\rm CO}$ & $x(e)$ & $\zeta$ [s$^{-1}$] \\
\hline 
\multicolumn{4}{l}{$V_{\rm LSR}=-11$ km~s$^{-1}$} \\ \hline
Caselli (2002) & $2.6\times10^5$ & & $\geq(2.91\pm0.91)\times10^{-9}$ \\
\\
Caselli (1998) & $2.6\times 10^5$ & $3\times10^{-5}$ & $(1.46\pm0.25)\times10^{-5}$ & $(5.48\pm2.55)\times10^{-13}$ \\\\
Luo (2024) & $1.0\times10^4$ & $8\times10^{-4}$ & $(2.58\pm0.08)\times10^{-7}$ & $(1.75\pm0.15)\times10^{-15}$ \\
& $2.6\times10^5$ & $3\times10^{-5}$ & $(1.18\pm0.03)\times10^{-8}$ & $(2.40\pm0.21)\times10^{-17}$ \\
& $1.0\times10^6$ & $8\times10^{-6}$ & $(4.79\pm0.13)\times10^{-9}$ & $(1.38\pm0.11)\times10^{-17}$ \\
\\
& $1.0\times10^4$ & $1\times10^{-4}$ & $(1.32\pm0.07)\times10^{-7}$ & $(8.10\pm2.02)\times10^{-17}$ \\
& $2.6\times10^5$ & $1\times10^{-4}$ & $(1.55\pm0.04)\times10^{-8}$ & $(6.84\pm0.48)\times10^{-17}$ \\
& $1.0\times10^6$ & $1\times10^{-4}$ & $(7.46\pm0.20)\times10^{-9}$ & $(6.75\pm0.42)\times10^{-17}$ \\
\hline
\multicolumn{4}{l}{$V_{\rm LSR}=-5$ km~s$^{-1}$} \\ \hline
Luo (2024) & $1.0\times10^7$ & $1\times10^{-4}$ & $\leq2.55\times10^{-10}$ & $\leq9.16\times10^{-19}$ \\
& $1.0\times10^8$ & $1\times10^{-4}$ & $\leq7.66\times10^{-11}$ & $\leq9.15\times10^{-19}$ \\
\hline
\end{tabular}
\tablefoot{$n(\ce{H2})=2.6\times10^5$ cm$^{-3}$ is the volume density of \ce{H2} constrained from the RADEX grid search, see Fig. \ref{fig:grid_search_methanol_11_E}. The first group of \cite{Luo_2024} used $f_{\ce{CO}}$ calculated from the measured $N(\ce{CO})=1.32\times10^{19}$ cm$^{-2}$ (only gas phase, inferred from \ce{C^18O}), whereas the second group used the 
$f_{\ce{CO}}$ (both gas and ice phase). }
\end{table*}

\section{Discussion} \label{sec:discussion}
\subsection{Cosmic ray ionisation rate}
\label{subsec:discussion:crir}
\begin{figure}
    \centering
    \includegraphics[width=\linewidth]{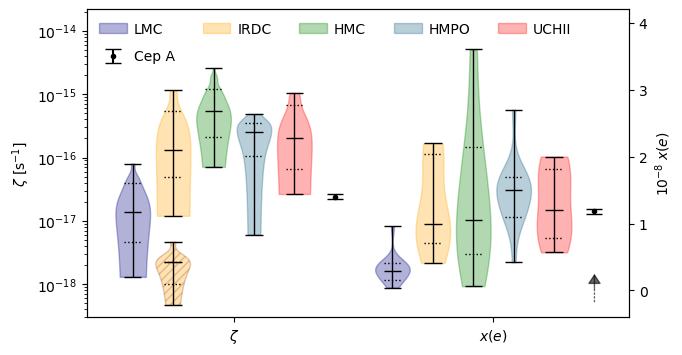}
    \caption{CRIRs (left) and electron fractions (right) for Cep A HW2 (at $V_{\rm LSR}=-11$ km~s$^{-1}$) compared to LMCs, IRDCs, HMCs, HMPOs and UCHIIs. Medians, minimum and maximum values are shown with black lines; 16\% and 84\% percentiles are shown with dotted black lines. The CRIR measurements for HMCs, HMPOs and UCHIIs are from \cite{Luo_2024}. The measurements for Cep A HW2 were calculated using the method from \cite{Luo_2024}. The CRIR measurements for IRDCs are from \cite{Miettinen_2011} (not hatched) and \cite[][the XR, case 2 in their Table 5]{Entekhabi2022} (hatched). The measurements for LMCs are from \cite{Redaelli_2025}. The CRIRs for LMCs and IRDCs were calculated using the method from \cite{Caselli_1998}.  All measurements are from single dish observations. The arrow shows the lower limit on the electron fraction calculated cf. \cite{Caselli_2002}. }
    \label{fig:crir_xe_comparison}
\end{figure}

Figure~\ref{fig:crir_xe_comparison} shows our estimates of the CRIR and electron fraction towards Cep A HW2 compared to infrared dark clouds (IRDCs), high-mass cores (HMCs), high-mass protostellar objects (HMPOs) and ultra compact HII regions (UCHIIs), which are all  precursors to high-mass stars at different evolutionary stages, as well as low-mass cores (LMCs), which are precursors to low-mass stars. The CRIRs for HMCs, HMPOs, UCHIIs and Cep A HW2 were calculated using the method from \cite{Luo_2024}, assuming $f(\ce{O})=2.51\times10^{-4}$, $f(\ce{N2})=3.8\times10^{-5}$ and $f(\ce{CO})=1\times10^{-4}$. The CRIR towards Cep A HW2 at $-11$~km~s$^{-1}$ is relatively low ($\sim$ a few 10$^{-17}$~s$^{-1}$, see Table~\ref{tab:X(e)_and_CRIR}), but it is consistent with other high-mass sources. The statistical error (from the column density measurements) is small, but the systematic errors associated with the fractional abundances of \ce{O}, \ce{N2} and \ce{CO} could be large. As a result, the absolute values may be incorrect, but any relative trend should indicate a true trend because the CRIRs are calculated in the same way. Furthermore, the ionisation rate measured using the method from \cite{Luo_2024} is not restricted to cosmic rays, but includes anything (e.g., x-rays) that initiates the same chemistry. Besides that, a recent study of \citet{Indriolo2026} who measured CRIR directly through \ce{H3^+} observations has shown that CRIR may be consistent within one cloud and vary from cloud to cloud as close as $\sim$100~pc.

Although the CRIR at $-11$ km~s$^{-1}$ is not generally high, it is almost two orders of magnitude higher than the upper limit of the CRIR at $-5$ km~s$^{-1}$ ($\sim10^{-18}$~s$^{-1}$). This suggests that there is a local enhancement of the CRIR at $-11$ km~s$^{-1}$ compared to $-5$ km~s$^{-1}$. Interestingly, most of the emission from COMs and ions is also associated with this kinematic component, see Section~\ref{subsec:discussion:abundance_coms} for further discussion hereof. The low value of the upper limit of the CRIR at $-5$ km~s$^{-1}$ is consistent with IRDCs \citep[e.g., ][]{Entekhabi2022}

The \cite{Luo_2024} method probes the CRIR from column density measurements of \ce{N2H+} and \ce{HCO+}, which we inferred from \ce{HC^18O+}. Both \ce{N2H+} and \ce{HC^18O+} exhibited an additional splitting of the velocity component at $-11$ km~s$^{-1}$, in fact, all ions and simple molecules with a good S/N exhibited this splitting, see Fig.~\ref{fig:hyperfine_fits}. \ce{C^17O} and \ce{NH3} are the only exceptions, but the complex profile of \ce{C^17O} may be better modelled with two components. Except for \ce{^13CO}, the bluer component is fainter, which could indicate that the gas is expanding. Although the lines from COMs generally had a lower S/N, none of them exhibited an additional splitting of the component at $-11$ km~s$^{-1}$. One explanation could be that COMs are found in the central regions of the protostellar core surrounded by a shell of expanding gas. \cite{Hernandez_2026} mapped the integrated intensities for different molecules in the Horsehead Nebula, showing that \ce{HC3N} was compact and centred around the dense core. On the other hand, COMs like \ce{CH3CN}, \ce{CH3CHO} and \ce{H2CCO} were generally more extended and centred behind the photodissociation region. The excitation temperatures and optical depths were poorly constrained by the fits with two components, so we estimated the column densities towards Cep~A using $\sigma$ from fits with only one component where the optical depth was not negligible. The intensities were integrated numerically, so the integrated intensities should not be affected hereby, but the widths would be smaller. We do not believe that the splitting of the component at $-11$ km~s$^{-1}$ has a large effect on our estimate of the CRIR towards Cep A HW2. 

The CRIRs for LMCs and IRDCs (Fig.~\ref{fig:crir_xe_comparison}) were calculated using the method from \cite{Caselli_1998}. They should be compared to the CRIR towards Cep A HW2 that was also calculated using the \cite{Caselli_1998} method ($\sim5.4\times10^{-13}$~s$^{-1}$). Thus, the CRIR is lowest in LMCs, higher in IRDCs and even higher towards Cep A HW2. In low-mass star-forming regions (LMSFRs), CRs are primarily produced by external sources such as supernovae, but in HMSFRs, CRs are also produced internally in protostellar objects by strong jets \citep[e.g.][]{Padovani2016}. Thus, the CRIR is expected to be higher in HMSFRs than LMSFRs, as observed. The CRIR towards Cep A HW2 calculated using the \cite{Caselli_1998} method is, however, unrealistically high, especially given the relatively low kinetic temperature in the region. Following \cite{Caselli_1998}, \cite{Redaelli_2024} also measured CRIRs that were overestimated by $2-4$ orders of magnitude compared to synthetic and real observations. \cite{Redaelli_2024} attributed the discrepancy to the neglect of doubly and triply deuterated forms of \ce{H3+}, and they found that the discrepancy was larger when deuteration is low as in our case where $R_D=(2.2\pm0.4)\times10^{-3}$. 

The electron fraction towards Cep A HW2 is typical compared to IRDCs, HMCs, HMPOs and UCHIIs; it appears to be the same in all of the included HMSFRs. The electron fraction in LMCs is slightly lower. 

\subsection{Abundance of COMs} 
\label{subsec:discussion:abundance_coms}
The fractional abundances were calculated as $N/N(\ce{H2})$ with $N(\ce{H2})$ adopted from the RADEX modelling of methanol (see Sect. \ref{subsec:radex_grid_search}) and are summarised in Table \ref{tab:column_density_abundance}. As argued in Section~\ref{subsec:discussion:crir}, the CRIR is locally enhanced at $-11$~km~s$^{-1}$ compared to $-5$~km~s$^{-1}$, and the component at $-11$~km~s$^{-1}$ is associated with most of the emission from COMs in the region. The abundance of methanol is higher by a factor of $\sim10^3$, and the abundance of methyl cyanide is higher by a factor of $\sim67$, assuming both components have the same source size. This increase in the abundance of COMs could be related to the increase in the CRIR. Indeed, laboratory experiments have shown that irradiating ices with CRs creates electrons that ionise molecules, which in turn recombine to produce radicals \citep{Gutierrez_2021}. These radicals are unstable and combine to form more complex molecules, including COMs \citep{Odberg_2016,Gutierrez_2021,Gaches_2026}. \citet{Shingledecker2018} in their modelling show that, for example, the highest abundance of methyl formate in TMC1 ($\sim10^{-10}$, close to our 3.78$\times10^{-10}$ wrt H) can be reached earlier during the cloud evolution with higher CRIR ($\sim10^{-16}-10^{-15}$~s$^{-1}$) and be an order of magnitude higher than with `standard' CRIR of $10^{-17}$~s$^{-1}$. However, with further increase to $10^{-14}$~s$^{-1}$, the abundance drops drastically (see panel $d$ of their Fig.~8). The increase of the abundances could, however, also be affected by other environmental factors. For example, the rotational temperature of methyl cyanide is a factor of 2 higher at $-11$ km~s$^{-1}$, but the kinetic temperatures of the two components (constrained from RADEX grid searches with low-energy methanol lines) are similar. The emission from low-energy methanol lines is stronger at $-11$ km~s$^{-1}$, whereas the emission from high-energy methanol lines is stronger at $-5$ km~s$^{-1}$. If the temperature is similar, the higher-energy transitions may be preferred at $-5$~km~s$^{-1}$ because this component is denser, see Table~\ref{tab:radex_results}. If, instead, there is a radiative source for high-energy excitation, it has to be local. However, such a source would probably increase the CRIR at $-5$~km~s$^{-1}$, contrary to our observations. \cite{Brogan2007} reported emission from COMs like \ce{CH3OCH3}, \ce{HCOOH} and \ce{NH2CHO} at $-5$ km~s$^{-1}$, which we did not detect, but from fewer COMs at $-11$ km~s$^{-1}$. \cite{Brogan2007} also reported emission from \ce{C^34S} (which we did not detect either), which is normally a very bright line, suggesting that the source of emission at $-5$ km~s$^{-1}$ is very compact. 

\begin{figure*}
    \centering
    \includegraphics[width=0.9\linewidth]{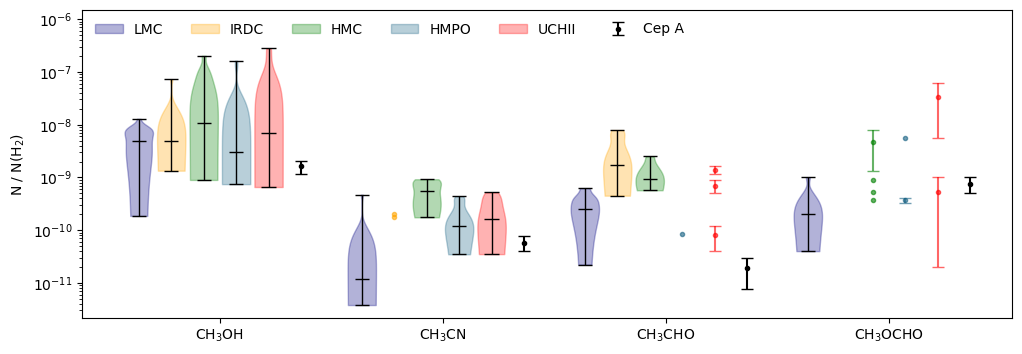}
    \caption{Abundance of COMs for Cep A HW2 (at $V_{\rm LSR}=-11$ km~s$^{-1}$) compared to LMCs, IRDCs, HMCs, HMPOs and UCHIIs. Medians, minimum and maximum values are shown with black lines. If fewer than 5 measurements are reported, the individual measurements are shown instead of a violin plot. The measurements for IRDCs, HMCs, HMPOs and UCHIIs are from \cite{Nummelin_1998,Ikeda_2001,Gerner_2014,Vasyunia_2014}. The measurements for LMCs are from \cite{Bacmann_2012,Bacmann_2016,Jimenez_Serra_2016,Chacon_Tanarro_2019,Nagy_2019,Lattanzi_2020,Jimenez_Serra_2021,Cabezas_2021,Scibelli_2021,Agundez_2021,Megias_2023,Scibelli_2024,Cabezas_2025}. All measurements are from single dish observations. }
    \label{fig:com_abundance_comparison}
\end{figure*}

Figure~\ref{fig:com_abundance_comparison} shows the fractional abundance of methanol, methyl cyanide, acetaldehyde and methyl formate towards Cep A HW2 compared to LMCs, IRDCs, HMCs, HMPOs and UCHIIs. Generally, the abundance of COMs seems lower in LMCs compared to IRDCs, HMCs, HMPOs and UCHIIs, but there is no clear trend as the high-mass cores evolve from starless clouds to protostellar cores with compact HII regions. This possibly indicates that the formation and/or desorption process for COMs is different in LMSFRs than in HMSFRs, where the formation and/or desorption of COMs is more efficient, possibly because of higher ionisation. The abundances of methanol, methyl cyanide and methyl formate towards Cep A HW2 are comparable to other HMSFRs, but the abundance of acetaldehyde is lower.  However, the measurements of acetaldehyde and methyl formate are scarce, so the distributions shown in Fig.~\ref{fig:com_abundance_comparison} may not necessarily reflect the general population of HMSFRs. 

\subsection{Chemical model}
We have run a set of 0D chemical models using the gas-grain code {\sc Nautilus} \citep{Ruaud2016, Wakelam2024} to examine the viability of our estimations of the density, gas temperature, and CRIR. Our models used the KIDA 2024 gas and grain networks \citep{Wakelam2024} and the default grain parameters for the models. The models use the estimated $n(\ce{H2}) = 2.6\times10^5$~cm$^{-3}$ and  $T_{\rm gas}$~= 33~K. Since Cep A is considered a massive star-forming region, we used an elevated UV field of 10 in the unit of the Draine field \citep{Draine1978}. We applied the default elemental composition which is EA1 `low-metals'  for all elements except He, \ce{C^+}, N, and O, taken from EA2 `high-metals' \citep[see][]{Wakelam2008}. We ran models with an $A_V$ of 5 and 30 magnitudes, and found that our model results did not appreciably change. Therefore, we present only the results for $A_V = 30$ mag. We ran a series of models with $\zeta = 10^{-18} - 10^{-13}$ s$^{-1}$, keeping all other parameters fixed. We do not aim to provide a best-fit analysis using these 0D models: while a best-fit may be achieved, since it is expected that the observed molecules trace different densities along the line of sight, we instead use the models as an extra viability check of the inferred parameters.

\begin{figure}
    \centering
    \includegraphics[width=0.9\columnwidth]{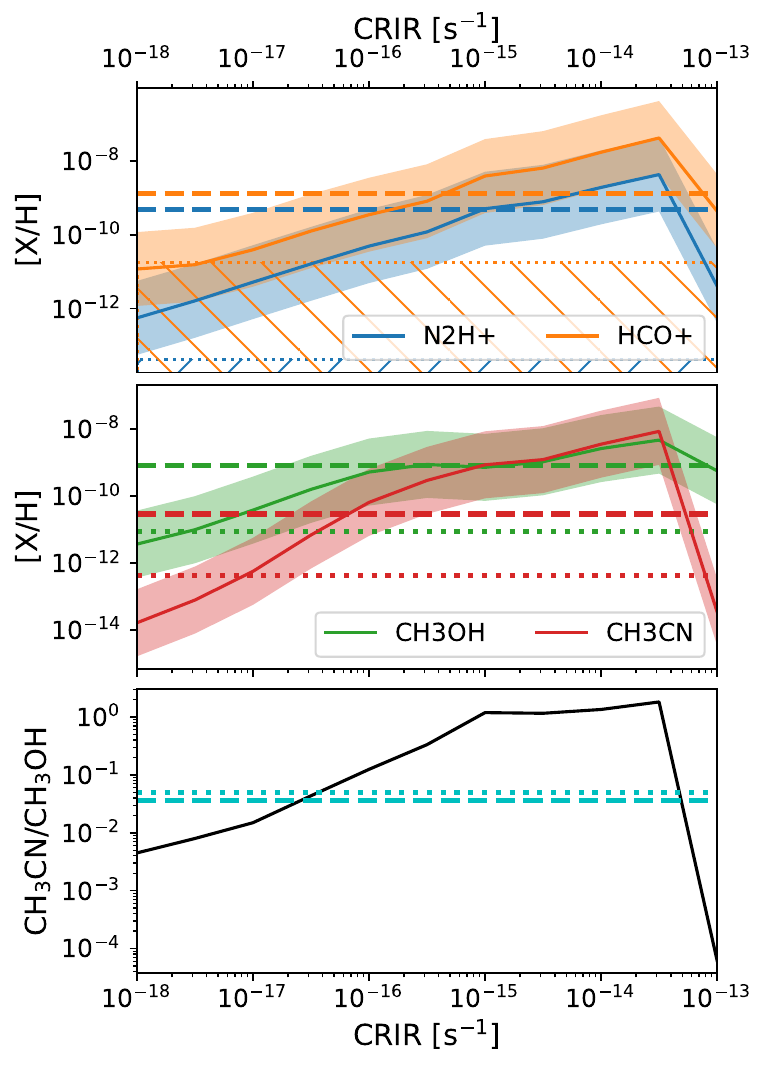}
    \caption{Top and middle: 0D Chemical model abundances for \ce{HCO+}, \ce{N2H+}, \ce{CH3OH}, and \ce{CH3CN} as a function of the \ce{H2} CRIR. The horizontal dashed lines indicate the observed inferred abundances. The solid lines with bands show the model results with a 1-dex spread. Bottom: Abundance ratio of \ce{CH3OH} and \ce{CH3CN} as a function of \ce{H2} CRIR. The horizontal dashed cyan line shows the observed abundance ratio, and the black line shows the model results.}
    \label{fig:chem_0D_model}
\end{figure}

The 0D models are run with fixed parameters as a function of time. We plot the averaged abundances from $t = 10^5 - 10^6$ years, since the age of Cep A is not known. Figure \ref{fig:chem_0D_model} shows the main results of key molecules as a function of the CRIR. In the figure, we show a band of 1 dex around the model results \citep[e.g.,][]{Vasyunin2004,Wakelam2005,Wakelam2010}. The dashed and dotted horizontal lines show the abundances for the -11~km~s$^{-1}$ and -5~km~s$^{-1}$ (the latter assuming $N(\ce{H2})=8\times10^{24}$~cm$^{-2}$) features, respectively, with the upper limits on \ce{HCO+} and \ce{N2H+} for the latter shown with hatches. We find that the model is able to reasonably reproduce the observed abundances and the ratio around the inferred CRIRs. In general, the abundances of \ce{N2H+} and \ce{HCO+} are under-estimated by the model. However, these molecules become abundant in less dense gas, and thus the inferred column density may be tracing a substantial component in the cloud envelope that is in our beam. Therefore, we would not expect a 0D chemical model to be a reasonable model for both dense tracers, such as \ce{CH3OH}, and lower-density species such as \ce{HCN}, and as such we only focus on the predicted COM abundances since these provide the tight constraints on density and temperature.

The abundance ratio of \ce{CH3CN}/\ce{CH3OH} is well-reproduced by an ionization rate $\approx 3\times 10^{-17} - 10^{-16}$~s$^{-1}$, in agreement with the observed inferred ratio for the -11 km s$^{-1}$ feature. Since the density and temperature were inferred using these COMs, we expect the ratio to be a more robust constraint. The abundances and abundance ratio are also reproduced by an extremely elevated ionization, $\zeta > 10^{-14}$~s$^{-1}$. However, at these densities, such an elevated CRIR heats the gas to temperatures exceeding 30 K, to over 100 K for ionisation rates $\zeta > 10^{-13}$~s$^{-1}$ \citep{Gaches_2026}. The chemical models provide further support for an inferred CRIR of $\zeta \approx (3-10)\times 10^{-17}$~s$^{-1}$. We note that the abundance ratio is also reproduced for the -5 km s$^{-1}$ for ionization rates $\approx 10^{-17}$ s$^{-1}$, an order of magnitude higher than the estimated values. However, at very low cosmic-ray ionization rates, there is less of an impact on the cosmic-ray ionization rate on the ice chemistry, so the ratio and the COM abundances may not probe these very low ionization rates.
 
\subsection{Deuteration}
\begin{figure*}
    \centering
    \includegraphics[width=0.9\linewidth]{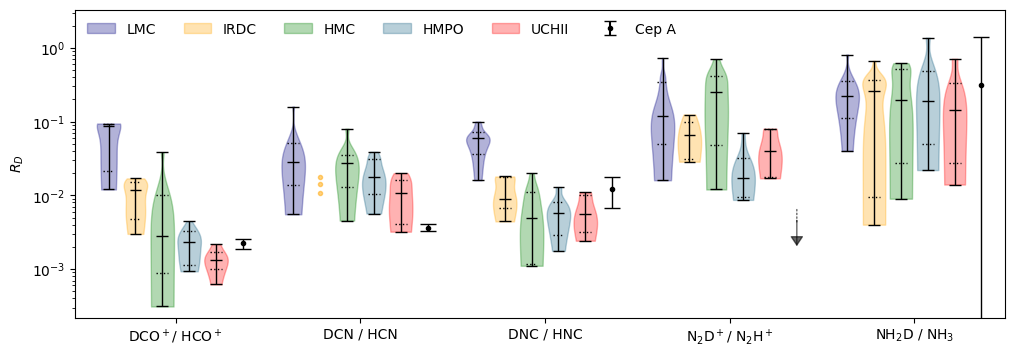}
    \caption{Deuterium fractions for Cep A HW2 (at $V_{\rm LSR}=-11$ km~s$^{-1}$) compared to LMCs, IRDCs, HMCs, HMPOs and UCHIIs. Medians, minimum and maximum values are shown with black lines. The measurements for IRDCs, HMCs, HMPOs and UCHIIs are from \cite{Pillai2007,Miettinen_2011,Fontani_2011,Fontani2014,Fontani2015,Gerner_2015}. The measurements for LMCs are from \cite{Hatchell2003,Roueff_2003,Roueff2005,Crapsi_2005,Crapsi2007,Punanova_2016,Harju_2017,Redaelli_2019,Giers_2023,Petrashkevich_2024,Petrashkevich_2026,Tasa-Chaveli_2025}. 
    If fewer than 5 measurements are reported, the individual measurements are shown instead of a violin plot. All measurements are from single dish observations. 
    The black arrow shows the upper limit of $R_D^{\ce{N2D+}}$ for Cep A HW2.  }
    \label{fig:deuterium_fraction_comparison}
\end{figure*}

Figure~\ref{fig:deuterium_fraction_comparison} shows the deuterium fractions $R_D^{\ce{DCO+}}$, $R_D^{\ce{DCN}}$, $R_D^{\ce{DNC}}$, $R_D^{\ce{N2D+}}$ and $R_D^{\ce{NH2D}}$ towards Cep A HW2 compared to LMCs, IRDCs, HMCs, HMPOs and UCHIIs. In general, LMCs appear to be highly deuterated, and $R_D^{\ce{DCO+}}$ seems to decrease as the high-mass protostar evolves. \cite{Kong_2015} and \cite{Caselli_2008} predicted from simulations that the deuterium fractions $R_D^{\ce{N2D+}}$ and $R_D^{\ce{H2D+}}$ would decrease when the CRIR increases because the electron fraction would increase, resulting in a larger dissociative recombination rate. The CRIRs used in the simulations were realistic, but in reality it is difficult to isolate a single parameter. For example, the CRIR is correlated with the temperature of the source which also changes as the protostar evolves. \cite{Kong_2015} found that the deuterium fraction decreases when the temperature is increased, which enhances the effect from the CRIR, but \cite{Caselli_2008} found that the effect of the CRIR on the deuterium fraction is most significant at temperatures below 15~K. In addition, \cite{Harju_2024} showed that the electron fraction increases more slowly than the CRIR, so the dissociation reactions may be outcompeted by other reactions or mechanisms. If we only compare our measurement towards Cep A HW2 with other HMPOs (which have CRIRs in the range $5\times10^{-18}-10^{-16}$~s$^{-1}$), $R_D^{\ce{DCO+}}$ does not appear to be correlated with the CRIR ($R_D^{\ce{DCN}}$ may be weakly correlated with the CRIR). On the other hand, \cite{Socci_2024} found that $R_D^{\ce{DCO+}}$ is indeed negatively correlated to the CRIR across Orion Molecular Clouds 2 and 3, as predicted by \cite{Caselli_2008} and \cite{Kong_2015}. Similar results were found for distant IRDCs \citep[AG351, AG354;][]{Sabatini2023} and a close Bok globule \citep[B335;][]{Cabedo2023}. The observed decrease of $R_D^{\ce{DCO^+}}$ with $\zeta$ may in fact be ruled by increasing temperature and thus the o/p ratio of \ce{H2}. Our results showed higher temperatures for Cep~A HW2, $T>20$~K; \citet{Shingledecker2016}, in the model of TMC-1, for $T=24$~K, showed that $R_D^{\ce{DCO^+}}$ should increase with CRIR up to about $10^{-16}$~s$^{-1}$ and again decrease with $\zeta>10^{-16}$~s$^{-1}$ (see panel $a$ of their Fig.~5). In Fig.~\ref{fig:deuterium_fraction_comparison}, $R_D^{\ce{DCN}}$ may also decrease as the protostar evolves, but the trend is much less significant. $R_D^{\ce{DNC}}$ reaches a minimum for HMCs, after which it stabilises or increases. The stable $R_D^{\ce{DNC}}$ is consistent with the model of \citet{Fontani2014}.

The \cite{Luo_2024} method for estimating the CRIR assumes that deuteration and \ce{CO} depletion is negligible. As a result, this method does not work for LMCs and IRDCs where deuteration and depletion is non-negligible. Fig.~\ref{fig:deuterium_fraction_comparison} shows that $R_D^{\ce{DCO+}}$ and $R_D^{\ce{N2H+}}$ (i.e., the deuterium fractions for the ions that were used to calculate the CRIR) are lower towards Cep A HW2 than in LMCs and IRDCs, as assumed. On the other hand, $R_D^{\ce{DNC}}$ towards Cep A HW2 is comparable to $R_D^{\ce{DNC}}$ in LMCs and IRDCs, suggesting that deuteration may not in general be negligible in HMSFRs. Indeed, $R_D^{\ce{DCO+}}$ is lower in HMCs, HMPOs and UCHIIs than in LMCs and IRDCs, whereas $R_D^{\ce{N2H+}}$ in HMCs, HMPOs and UCHIIs is comparable to $R_D^{\ce{N2H+}}$ in LMCs and IRDCs. At $n(\ce{H2})=2.6\times10^5$ cm$^{-3}$, the CRIR decreased when using the measured $f_{\ce{CO}}$ (where some \ce{CO} is depleted) instead of the "universal" value, and at $n(\ce{H2})\sim10^4$ cm$^{-3}$, the CRIR increased (see Table \ref{tab:X(e)_and_CRIR}), so it is unclear what effect this has on our estimate of the CRIR. \cite{Redaelli_2024} find that for $R_D\leq10^{-2}$, the CRIR depends only weakly on the depletion factor.

\section{Conclusions} \label{sec:conclusion}
We analysed high-resolution spectra from a single dish observation in the 3-4 mm and 7 mm bands towards the nearby, massive protostar Cep A HW2. We did a blind inventory of its chemical composition and estimated its CRIR through analytical chemistry. Our main findings are as follows.

\begin{enumerate}
    \item Cep A HW2 has two kinematic components. COMs, ions and simple neutrals are primarily emitted at $V_{\rm LSR}=(-10.56\pm0.06)$ km~s$^{-1}$, but methanol and methyl cyanide are also emitted at $(-4.73\pm0.09)$ km~s$^{-1}$. Some ions like \ce{DCO+} and \ce{N2H+} exhibit an additional splitting of the component at $-11$ km~s$^{-1}$.  
    \item The fractional abundance of (low energy) \ce{CH3OH} constrained from a RADEX grid search is $(1.6\pm0.5)\times10^{-9}$ at $-11$ km~s$^{-1}$. The fractional abundances of \ce{CH3CN}, \ce{t-HCOOH}, \ce{H2CCO}, \ce{CH3CHO} and \ce{CH3OCHO} are $(5.9\pm1.9)\times10^{-11}$, $(8\pm14)\times10^{-11}$, $(1.7\pm2.0)\times10^{-11}$, $(1.9\pm1.1)\times10^{-11}$ and $(8\pm3)\times10^{-10}$, respectively. $3\sigma$ upper limits on the fractional abundances of \ce{CH3OCH3} and \ce{NH2CHO} are $\leq1.6\times10^{-9}$ and $\leq3.0\times10^{-12}$, respectively. The abundances are generally consistent with other HMSFRs.  
    \item The deuterium fractions for \ce{DCO+}/\ce{HCO+}, DNC/HNC, DCN/HCN and \ce{NH2D}/\ce{NH3} are in the range $0.002-0.3$ and agree well with other HMPOs or UCHIIs. 
    \item Following the method proposed by \cite{Luo_2024}, the CRIR is $(6.8\pm0.5)\times10^{-17}$~s$^{-1}$ at $-11$ km~s$^{-1}$ and $\leq9.2\times10^{-19}$~s$^{-1}$ at $-5$ km~s$^{-1}$. The electron fraction at $-11$~km~s$^{-1}$ is typical compared to IRDCs, HMCs, HMPOs and UCHIIs. 
    \item Although the CRIRs are not generally elevated, our results show that the CRIR of the kinematic component associated with most of the COM's emission in the region is locally enhanced. This suggests that a higher CRIR can increase the gas phase abundance of COMs in HMSFRs too. We did not detect any deuterated species (nor their hydrogenated counterparts) for the component with the low CRIR, and deuteration (particularly of \ce{HCO+}) seems to decrease as the high-mass protostar evolves.
\end{enumerate}

\begin{acknowledgements}
EWN acknowledges support from the Chalmers Astrophysics and Space Sciences Summer (CASSUM) program.      
\end{acknowledgements}

\bibliographystyle{aa}
\bibliography{CepA}

\appendix
\onecolumn

\section{Spectra}
\begin{figure*}
    \centering
    \includegraphics[width=0.9\linewidth]{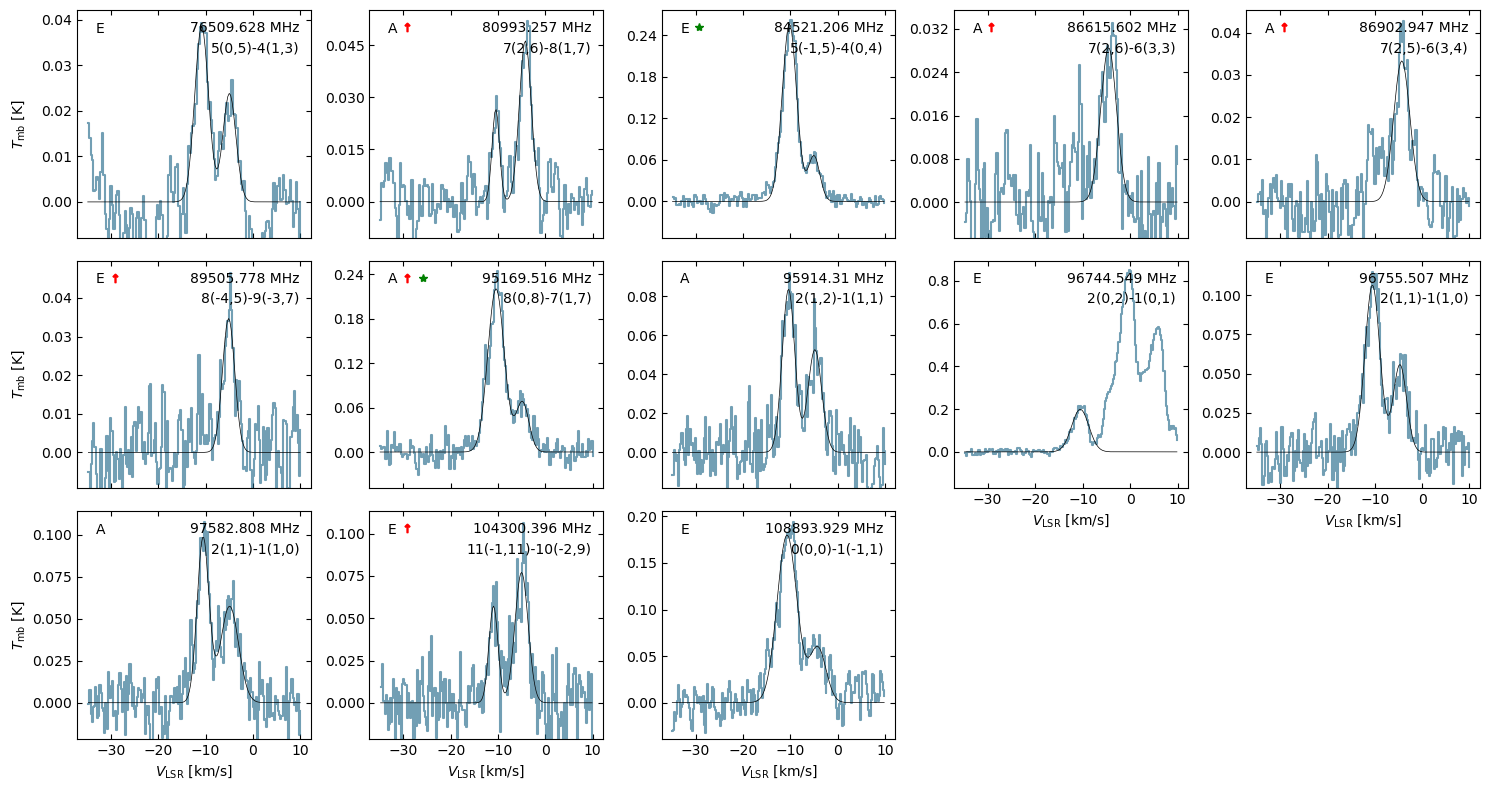}
    \caption{Spectra of the detected methanol (\ce{CH3OH}) lines towards Cep~A HW2. The blue lines show the baseline-subtracted spectra, and the black lines show the fits. The "A"/"E" in the upper left corners indicate methanol A/E, respectively. A red upward arrow indicates that $E_u>50$~K, and a green star indicates a maser. The frequency and quantum numbers of the transition are listed in the upper right corners. }
    \label{fig:methanol_peaks}
\end{figure*}

\begin{figure*}
    \centering
    \includegraphics[width=0.9\linewidth]{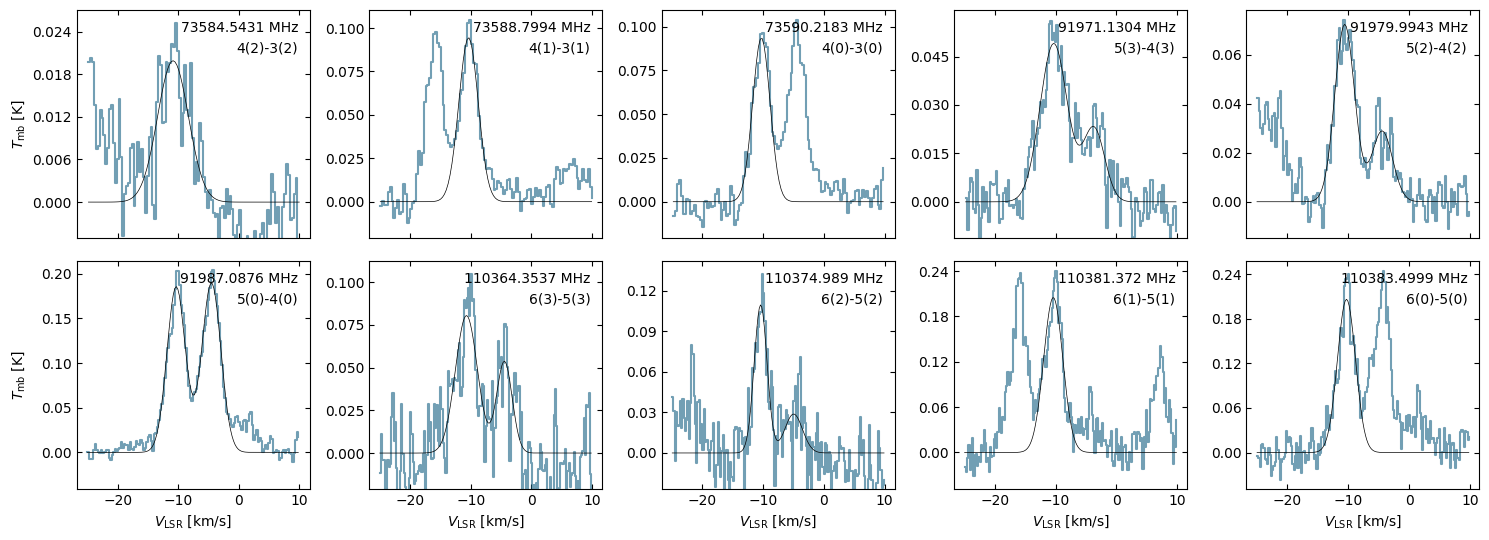}
    \caption{Spectra of the detected methyl cyanide (\ce{CH3CN}, blue) lines and their fits (black) towards Cep~A HW2.}
    \label{fig:methyl_cyanide_peaks}
\end{figure*}

\begin{figure*}
    \centering
    \includegraphics[width=0.9\linewidth]{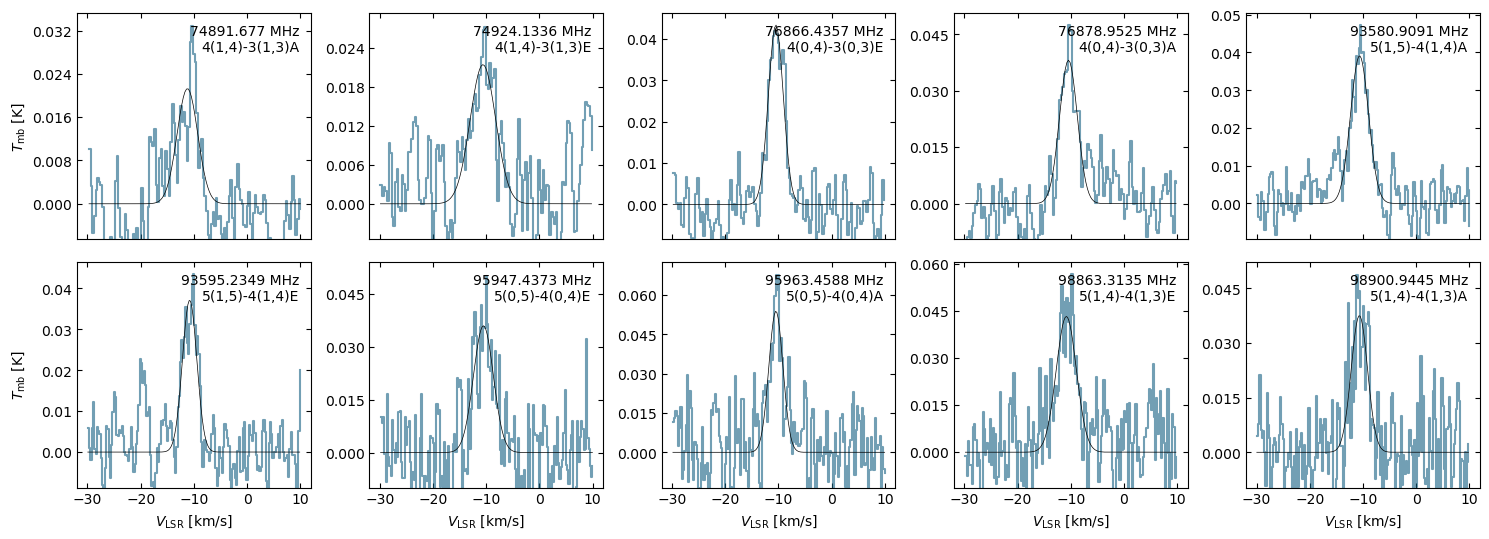}
    \caption{Spectra of the detected acetaldehyde (\ce{CH3CHO}, blue) lines and their fits (black) towards Cep~A HW2. }
    \label{fig:acetaldehyde_peaks}
\end{figure*}

\begin{figure*}
    \centering
    \includegraphics[width=0.9\linewidth]{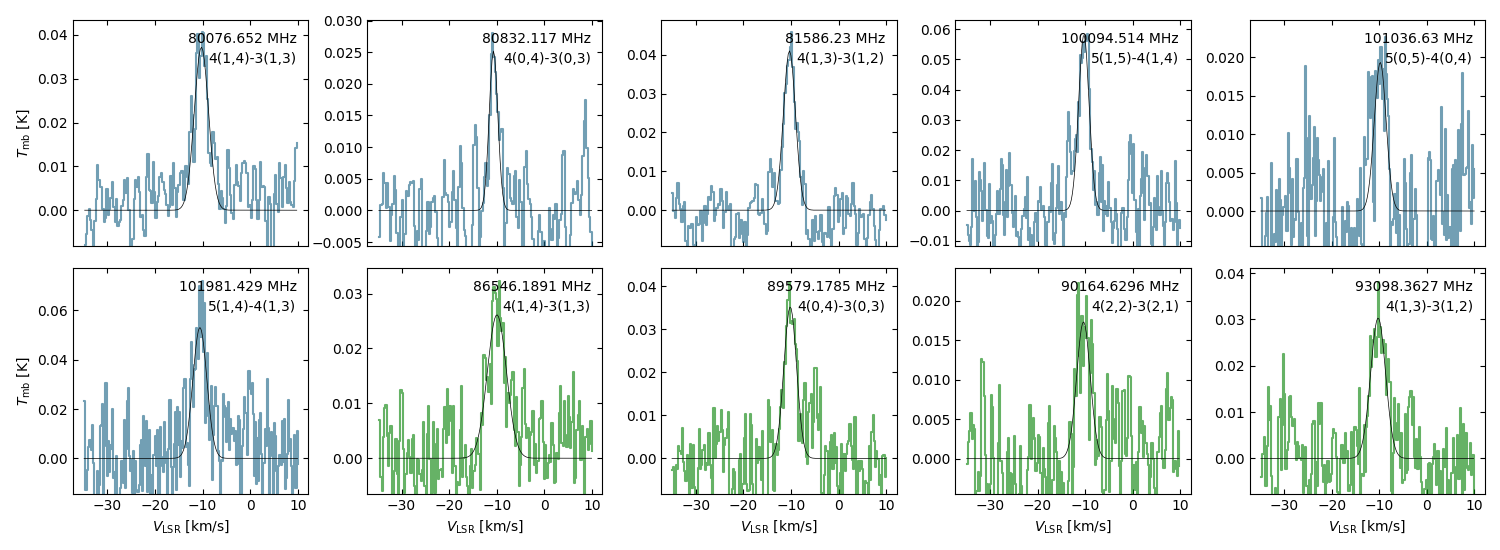}
    \caption{Spectra of the detected ketene (\ce{H2CCO}, first six plots in blue) and trans-formic acid (t-HCOOH, last four plots in green) lines  and their fits (black) towards Cep~A HW2. }
    \label{fig:ketene_peaks}
\end{figure*}

\begin{figure*}
    \centering
    \includegraphics[width=.72\linewidth]{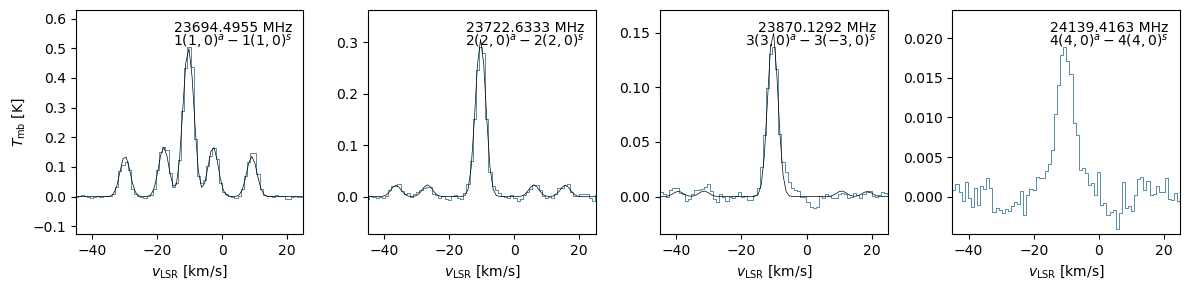}
    \caption{Spectra of the detected ammonia (\ce{NH3}, blue) lines and their fits (black) towards Cep~A HW2. Only the $1-1$, $2-2$ and $3-3$ transitions were used to constrain the column density and kinetic temperature.}
    \label{fig:ammonia_peaks}
\end{figure*}

\begin{figure*}
    \centering
    \includegraphics[width=0.9\linewidth]{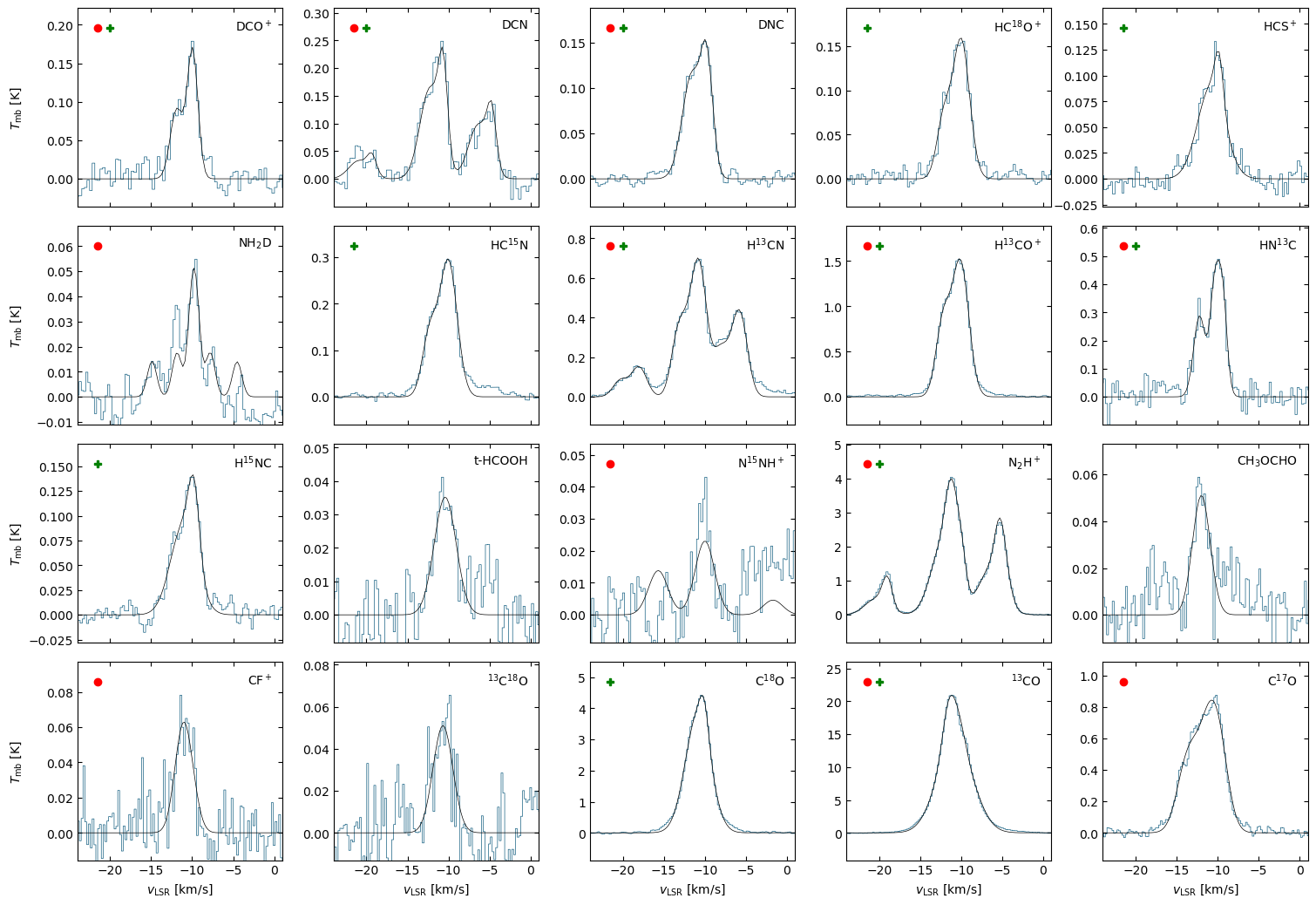}
    \caption{Spectra of the detected ions, simple neutrals, methyl formate (\ce{CH3OCHO}) and trans-formic acid (\ce{t-HCOOH}, 89579.1785~MHz) in blue and their fits (black) towards Cep~A HW2. A red dot indicates that the transition is fitted with an HFS, and a green plus indicates that the transition was fitted with two velocity components. }
    \label{fig:hyperfine_fits}
\end{figure*}

\section{Additional Tables and Figures}
\begin{figure*}
    \centering
    \includegraphics[width=0.9\linewidth]{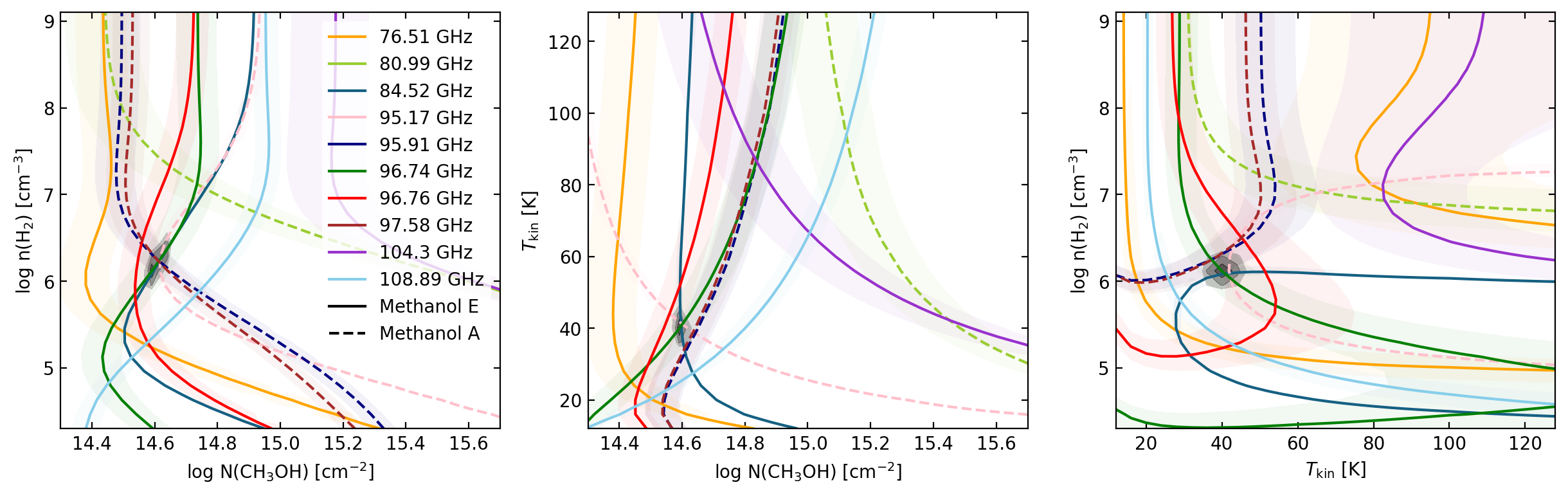}
    \includegraphics[width=0.9\linewidth]{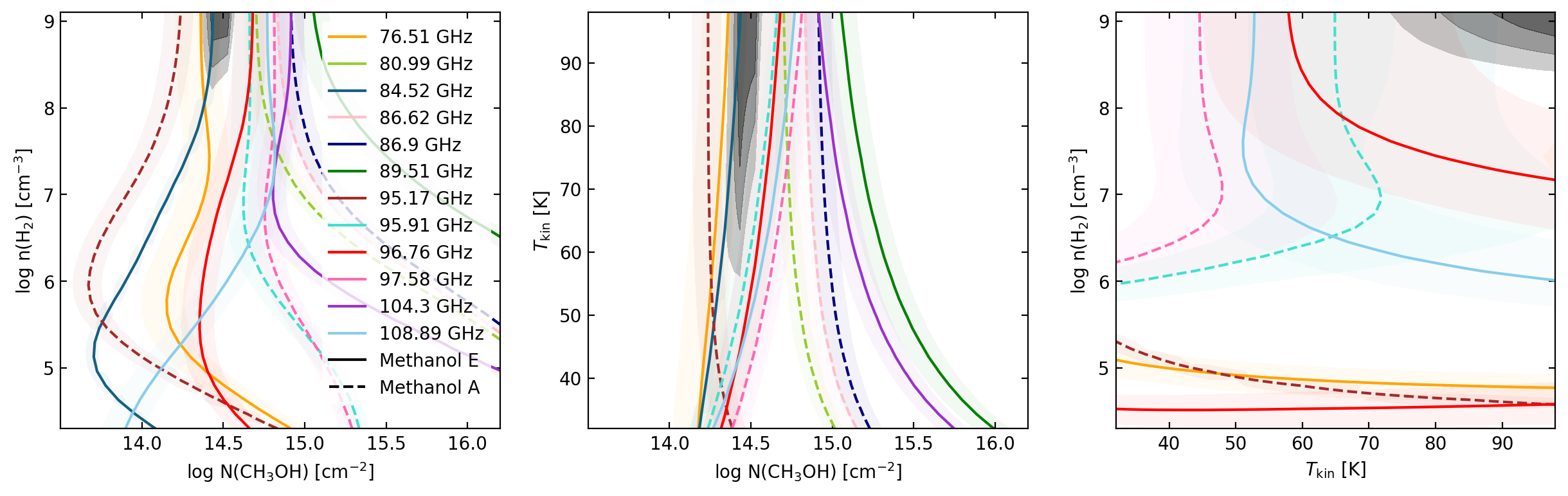}
    \caption{RADEX grid search using all methanol E and A lines at $-11$ km~s$^{-1}$ (top) and $-5$ km~s$^{-1}$ (bottom). In some of the plots, there are multiple lines corresponding to the emission lines at 76.51, 96.74 and 96.76 GHz because multiple excitation conditions could reproduce the observed intensities. }
    \label{fig:grid_search_methanol_all_lines}
\end{figure*}

\begin{table*}
    \caption{The velocities and line widths of the lines fitted with two component around $-11$~km~s$^{-1}$.}\label{tab:two-components-10}
    \centering
    \begin{tabular}{lrrrrr}
    \hline 
Species & $\nu$ [MHz] & $V_{\rm LSR,1}$ [km~s$^{-1}$] & $\sigma_1$ [km~s$^{-1}$] & $V_{\rm LSR,2}$ [km~s$^{-1}$] & $\sigma_2$ [km~s$^{-1}$] \\
\hline
\ce{DCO^+} & 72039.3031 & $-11.86\pm0.23$ & $0.33\pm0.19$ & $-9.91\pm0.11$ & $0.27\pm0.09$ \\
\ce{DCN} & 72414.6936 & $-11.25\pm0.35$ & $0.52\pm0.09$ & $-9.65\pm0.07$ & $0.24\pm0.06$ \\
\ce{DNC} & 76305.7270 & $-10.75\pm1.21$ & $0.50\pm0.17$ & $-8.93\pm0.48$ & $0.25\pm0.28$ \\
\ce{HC^{18}O^+} & 85162.2231 & $-12.14\pm0.82$ & $0.84\pm0.51$ & $-10.00\pm0.43$ & $1.03\pm0.29$ \\
\ce{HCS^+} & 85347.8900 & $-10.76\pm0.42$ & $1.67\pm0.30$ & $-9.80\pm0.32$ & $0.51\pm0.40$ \\
\ce{HC^{15}N} & 86054.9664 & $-12.31\pm0.24$ & $0.88\pm0.15$ & $-10.02\pm0.13$ & $1.07\pm0.09$ \\
\ce{H^{13}CN} & 86339.9214 & $-12.22\pm0.11$ & $0.37\pm0.03$ & $-9.94\pm0.06$ & $0.40\pm0.02$ \\
\ce{H^{13}CO^+} & 86754.2884 & $-12.21\pm0.09$ & $0.37\pm0.04$ & $-10.09\pm0.06$ & $0.43\pm0.02$ \\
\ce{HN^{13}C} & 87090.8500 & $-11.66\pm0.11$ & $0.30\pm0.13$ & $-9.37\pm0.07$ & $0.26\pm0.04$ \\
\ce{H^{15}NC} & 88865.7150 & $-11.03\pm0.24$ & $1.59\pm0.12$ & $-9.77\pm0.10$ & $0.65\pm0.13$ \\
\ce{N_2H^+} & 93173.3977 & $-11.09\pm0.11$ & $0.59\pm0.02$ & $-9.83\pm0.02$ & $0.24\pm0.02$ \\
\ce{C^{18}O} & 109782.1734 & $-10.85\pm0.02$ & $1.53\pm0.01$ & $-10.07\pm0.03$ & $0.53\pm0.04$ \\
\ce{^{13}CO} & 110201.3697 & $-10.18\pm0.04$ & $0.33\pm0.02$ & $-9.87\pm0.02$ & $0.85\pm0.00$ \\
\hline
\end{tabular}
\end{table*}

\begin{table*}
    \caption{Column densities and abundances obtained assuming LTE. The relative abundances were calculated assuming $N(\ce{CH3OH})=(3.52\pm0.08)\times10^{14}$~cm$^{-2}$ and $N(\ce{H2})=(2.2\pm0.6)\times10^{23}$~cm$^{-2}$ at $-11$~km~s$^{-1}$; $N(\ce{CH3OH})=(1.4\pm0.1)\times10^{14}$~cm$^{-2}$ and $N(\ce{H2})=8\times10^{24}$~cm$^{-2}$ at $-5$~km~s$^{-1}$ from the RADEX search.}
    \label{tab:column_density_abundance}
    \centering
    \begin{tabular}{lrrrrr}\hline
    \multicolumn{6}{c}{Rotational diagrams}\\
    \hline 
Species & $Q_\text{rot}$ & $T_\text{rot}$ [K] & $N$ [cm$^{-2}$] & $N/N(\ce{CH3OH})$ & $N/N(\ce{H2})$ \\
    \hline
\ce{CH3OH}-A \scriptsize{$-$11 km~s$^{-1}$} & $448\pm95$ & $37\pm5$ & $(1.04\pm0.26)\times10^{14}$ \\
\ce{CH3OH}-E \scriptsize{$-$11 km~s$^{-1}$} & $112\pm20$ & $17\pm2$ & $(9.96\pm2.27)\times10^{13}$ \\
\ce{CH3OH}-A \scriptsize{$-$5 km~s$^{-1}$} & $2950\pm1340$ & $113\pm29$ & $(3.59\pm1.78)\times10^{14}$ \\
\ce{CH3OH}-E \scriptsize{$-$5 km~s$^{-1}$} & $249\pm140$ & $26\pm8$ & $(5.41\pm3.65)\times10^{13}$ \\
\ce{CH3CN} \scriptsize{$-11$ km~s$^{-1}$} & $681\pm86$ & $49\pm4$ & $(1.28\pm0.17)\times10^{13}$ & $(3.65\pm0.49)\times10^{-2}$ & $(5.90\pm1.88)\times10^{-11}$ \\
\ce{CH3CN} \scriptsize{$-5$ km~s$^{-1}$} & $257\pm33$ & $26\pm2$ & $(6.88\pm0.98)\times10^{12}$ & $(4.91\pm0.79)\times10^{-2}$ & $\sim 8.3\times10^{-13}$ \\
\ce{t-HCOOH} & $309\pm514$ & $32\pm37$ & $(1.71\pm2.99)\times10^{13}$ & $(4.88\pm8.51)\times10^{-2}$ & $(7.88\pm13.95)\times10^{-11}$ \\
\ce{H2CCO} & $524\pm515$ & $40\pm27$ & $(1.37\pm1.44)\times10^{13}$ & $(5.61\pm5.97)\times10^{-2}$ & $(1.72\pm2.00)\times10^{-11}$ \\
\ce{CH3CHO}-A & $119\pm76$ & $14\pm6$ & $(2.12\pm1.62)\times10^{12}$ & $(6.04\pm4.61)\times10^{-3}$ & $(9.76\pm7.97)\times10^{-12}$ \\
\ce{CH3CHO}-E & $105\pm62$ & $13\pm5$ & $(1.98\pm1.44)\times10^{12}$ & $(5.64\pm4.09)\times10^{-3}$ & $(9.12\pm7.11)\times10^{-12}$ \\
\ce{CH3CHO} & $112\pm49$ & $13\pm4$ & $(4.10\pm2.16)\times10^{12}$ & $(1.17\pm0.61)\times10^{-2}$ & $(1.89\pm1.13)\times10^{-11}$ \\
\\
 \multicolumn{6}{c}{Optically thin estimates}\\
    \hline 
Species & $Q_{\rm rot}$ & $T_{\rm ex}$ [K] & $N$ [cm$^{-2}$] & $N/N(\ce{CH3OH})$ & $N/N(\ce{H2})$ \\
    \hline
\ce{t-HCOOH} & $328$ & $33$ & $(1.72 \pm 0.49)\times 10^{13}$ & & $(7.93 \pm 3.21)\times 10^{-11}$ \\ 
\ce{CH_3OCHO} & $4774$ & $33$ & $(1.65 \pm 0.26)\times 10^{14}$ & & $(7.56 \pm 2.50)\times 10^{-10}$ \\
\ce{CH3OCH3} & $8787$ & $33$ & $\leq4.38\times10^{14}$ & $\leq9.89\times10^{-1}$ & $\leq1.60\times10^{-9}$ \\
\ce{NH2CHO} & $1074$ & $33$ & $\leq4.37\times10^{12}$ & $\leq1.24\times10^{-2}$ & $\leq2.01\times10^{-11}$ \\
\hline
Species & $Q_{\rm rot}$ & $T_{\rm ex}$ [K] & $N$ [cm$^{-2}$] & $N$(main)\tablefootmark{a} [cm$^{-2}$] & $N/N(\ce{H2})$ \\
    \hline
\ce{DCO^+} & $3.01 \pm 0.04$ & $4.59$ & $(1.28 \pm 0.13)\times 10^{12}$ & & $(5.87 \pm 1.80)\times 10^{-12}$ \\ 
\ce{DCN} & $3.00 \pm 0.04$ & $4.59$ & $(5.40 \pm 0.43)\times 10^{12}$ & & $(2.48 \pm 0.74)\times 10^{-11}$ \\ 
\ce{HC^{18}O^+} \scriptsize{$-$11 km~s$^{-1}$} & $2.61 \pm 0.04$ & $4.59$ & $(1.10 \pm 0.15)\times 10^{12}$ & $(5.73\pm0.76)\times10^{14}$ & $(5.04 \pm 1.60)\times 10^{-12}$ \\ 
\ce{HC^{18}O^+} \scriptsize{$-$5 km~s$^{-1}$} & $2.61$ & $4.59$ & $\leq5.45\times 10^{11}$ & $\leq2.85\times10^{14}$ & $\leq6.82\times 10^{-14}$ \\ 
\ce{HCS^+} & $4.83 \pm 0.08$ & $4.59$ & $(4.49 \pm 0.92)\times 10^{12}$ & & $(2.07 \pm 0.73)\times 10^{-11}$ \\ 
\ce{HC^{15}N} & $2.59 \pm 0.04$ & $4.59$ & $(3.32 \pm 0.22)\times 10^{12}$ & $(1.48\pm0.10)\times10^{15}$ & $(1.52 \pm 0.45)\times 10^{-11}$ \\ 
\ce{H^{15}NC} & $2.52 \pm 0.04$ & $4.59$ & $(1.63 \pm 0.13)\times 10^{12}$ & $(7.24\pm0.58)\times10^{14}$ & $(7.47 \pm 2.24)\times 10^{-12}$ \\  
\ce{N^{15}NH^+} & $2.46 \pm 0.03$ & $4.59$ & $(4.66 \pm 1.11)\times 10^{11}$ & $(2.08\pm0.49)\times10^{14}$ & $(2.14 \pm 0.80)\times 10^{-12}$ \\ 
\ce{N_2H^+} \scriptsize{$-$11 km~s$^{-1}$} & $2.42 \pm 0.03$ & $4.59$ & $(5.14 \pm 0.29)\times 10^{13}$ & & $(2.36 \pm 0.70)\times 10^{-10}$ \\
\ce{N_2H^+} \scriptsize{$-$5 km~s$^{-1}$} & $2.42$ & $4.59$ & $\leq6.70\times 10^{11}$ & & $\leq8.39\times 10^{-14}$ \\
\ce{CF^+} & $2.24 \pm 0.03$ & $4.59$ & $(3.51 \pm 0.64)\times 10^{12}$ & & $(1.62 \pm 0.55)\times 10^{-11}$ \\ 
\ce{^{13}C^{18}O} & $2.20 \pm 0.03$ & $4.59$ & $(2.38 \pm 0.73)\times 10^{14}$ & $(8.64\pm2.66)\times10^{18}$ & $(1.09 \pm 0.46)\times 10^{-9}$ \\ 
\ce{C^{18}O} & $2.12 \pm 0.03$ & $4.59$ & $(2.53 \pm 0.15)\times 10^{16}$ & $(1.32\pm0.08)\times10^{19}$ & $(1.16 \pm 0.34)\times 10^{-7}$ \\ 
\ce{^{13}CO} & $2.11 \pm 0.03$ & $4.59$ & $(1.46 \pm 0.08)\times 10^{17}$ & $(1.01\pm0.06)\times10^{19}$ & $(6.70 \pm 1.98)\times 10^{-7}$ \\ 
\\
 \multicolumn{6}{c}{Optically thick estimates} \\ 
 \hline
Species & $Q_{\rm rot}$ & $T_{\rm ex}$ [K] & $N$ [cm$^{-2}$] & $N$(main)\tablefootmark{a} [cm$^{-2}$] & $N/N(\ce{H2})$ \\ \hline
 \ce{DNC} & $1.96 \pm 0.01$ & $2.89 \pm 0.01$ & $(1.08 \pm 0.49)\times 10^{13}$ & & $(4.95 \pm 2.67)\times 10^{-11}$ \\ 
\ce{NH_2D} & $2.79 \pm 0.00$ & $2.88 \pm 0.38$ & $(1.26 \pm 4.22)\times 10^{15}$ & & $(5.80 \pm 19.49)\times 10^{-9}$ \\ 
\ce{H^{13}CN} & $3.08 \pm 0.49$ & $5.63 \pm 1.03$ & $(3.53 \pm 1.83)\times 10^{12}$ & $(2.46\pm1.27)\times10^{14}$ & $(1.62 \pm 0.96)\times 10^{-11}$ \\ 
\ce{H^{13}CO^+} & $2.57 \pm 0.04$ & $4.59 \pm 0.08$ & $(5.19 \pm 0.61)\times 10^{12}$ & $(3.61\pm0.43)\times10^{14}$ & $(2.39 \pm 0.75)\times 10^{-11}$ \\ 
\ce{HN^{13}C} & $1.90 \pm 0.01$ & $3.17 \pm 0.02$ & $(1.26 \pm 0.04)\times 10^{13}$ & $(8.79\pm0.31)\times10^{14}$ & $(5.81 \pm 1.69)\times 10^{-11}$ \\ 
\ce{C^{17}O} & $2.27 \pm 0.21$ & $5.11 \pm 0.58$ & $(2.12 \pm 0.86)\times 10^{15}$ & $(3.55\pm1.45)\times10^{18}$ & $(9.76 \pm 4.87)\times 10^{-9}$ \\
\\
 \hline
Species & $Q_{\rm rot}$ & $T_{\rm rot}$ [K] & $T_{\rm ex}$ [K] & $N$ [cm$^{-2}$] & $N/N(\ce{H2})$ \\ \hline
\ce{NH3} & & $21.9\pm1.3$ & $3.9\pm0.6$ &  $(5.39\pm2.78)\times10^{15}$ & $(2.44\pm1.44)\times10^{-8}$\\ \hline
\end{tabular}
\tablefoot{\tablefoottext{a}{Inferred column density of the main isotopologues (CO, HNC, HCN, \ce{HCO^+}, \ce{N2H^+}). }}
\end{table*}

\begin{landscape}

\begin{center}
\begin{longtable}{lrrrrrrrrr}
\caption{The detected lines of COMs used in RADEX grid search and rotational diagrams and parameters of their Gaussian fits. }
\label{tab:detected_COM_lines} \\
\hline
Line & $\nu$ & $\theta_b$ & $A$ & $E_u$ & $W$ & $V_{\rm LSR}$ & $\sigma$ & $T_{\rm peak}$ & rms \\
& [MHz] & [$''$] & [$10^{-5}$ s$^{-1}$] & [K] & [K~km~s$^{-1}$] & [km~s$^{-1}$] & [km~s$^{-1}$] & [mK] & [mK] \\ \hline 
\ce{CH3OH}(4$_{-1,4}$--3$_{0,3}$)-E & 36169.29 & 103 & 0.015 & 29 & $0.54 \pm 0.04$ & $-9.59 \pm 0.07$ & $1.77 \pm 0.07$ & 122 &  7 \\
\ce{CH3OH}(5$_{0,5}$--4$_{1,3}$)-E & 76509.628 & 48 & 0.090 & 48 & $0.14 \pm 0.03$ & $-10.84 \pm 0.12$ & $1.46 \pm 0.12$ & 39 &  6 \\
\ce{CH3OH}(5$_{0,5}$--4$_{1,3}$)-E & 76509.628 & 48 & 0.090 & 48 & $0.08 \pm 0.02$ & $-4.98 \pm 0.19$ & $1.38 \pm 0.19$ & 24 &  6 \\
\ce{CH3OH}(7$_{2,6}$--8$_{1,7}$)-A & 80993.257 & 46 & 0.104 & 103 & $0.05 \pm 0.01$ & $-10.47 \pm 0.11$ & $0.78 \pm 0.11$ & 27 &  6 \\
\ce{CH3OH}(7$_{2,6}$--8$_{1,7}$)-A & 80993.257 & 46 & 0.104 & 103 & $0.15 \pm 0.02$ & $-4.23 \pm 0.08$ & $1.26 \pm 0.08$ & 46 &  6 \\
\ce{CH3OH}(5$_{-1,5}$--4$_{0,4}$)-E & 84521.206 & 44 & 0.197 & 40 & $0.98 \pm 0.03$ & $-10.21 \pm 0.02$ & $1.53 \pm 0.02$ & 256 &  6 \\
\ce{CH3OH}(5$_{-1,5}$--4$_{0,4}$)-E & 84521.206 & 44 & 0.197 & 40 & $0.20 \pm 0.02$ & $-5.03 \pm 0.06$ & $1.23 \pm 0.07$ & 66 &  6 \\
\ce{CH3OH}(7$_{2,6}$--6$_{3,3}$)-A & 86615.602 & 43 & 0.068 & 103 & $0.12 \pm 0.03$ & $-4.54 \pm 0.15$ & $1.66 \pm 0.15$ & 29 &  6 \\
\ce{CH3OH}(7$_{2,5}$--6$_{3,4}$)-A & 86902.947 & 43 & 0.069 & 103 & $0.14 \pm 0.03$ & $-4.31 \pm 0.12$ & $1.72 \pm 0.12$ & 33 &  6 \\
\ce{CH3OH}(8$_{-4,5}$--9$_{-3,7}$)-E & 89505.778 & 41 & 0.076 & 171 & $0.11 \pm 0.03$ & $-5.15 \pm 0.13$ & $1.30 \pm 0.13$ & 35 &  7 \\
\ce{CH3OH}(8$_{0,8}$--7$_{1,7}$)-A & 95169.516 & 39 & 0.426 & 84 & $0.96 \pm 0.06$ & $-10.42 \pm 0.04$ & $1.74 \pm 0.05$ & 220 &  12 \\
\ce{CH3OH}(8$_{0,8}$--7$_{1,7}$)-A & 95169.516 & 39 & 0.426 & 84 & $0.23 \pm 0.05$ & $-4.83 \pm 0.12$ & $1.37 \pm 0.13$ & 67 &  12 \\
\ce{CH3OH}(2$_{1,2}$--1$_{1,1}$)-A & 95914.31 & 39 & 0.249 & 21 & $0.19 \pm 0.04$ & $-4.84 \pm 0.13$ & $1.43 \pm 0.14$ & 52 &  11 \\
\ce{CH3OH}(2$_{1,2}$--1$_{1,1}$)-A & 95914.31 & 39 & 0.249 & 21 & $0.28 \pm 0.04$ & $-10.41 \pm 0.08$ & $1.33 \pm 0.08$ & 83 &  11 \\
\ce{CH3OH}(2$_{0,2}$--1$_{0,1}$)-E & 96744.549 & 38 & 0.341 & 20 & $0.68 \pm 0.10$ & $-10.54 \pm 0.08$ & $1.40 \pm 0.08$ & 194 &  26 \\
\ce{CH3OH}(2$_{1,1}$--1$_{1,0}$)-E & 96755.507 & 38 & 0.262 & 28 & $0.42 \pm 0.05$ & $-10.51 \pm 0.07$ & $1.58 \pm 0.08$ & 106 &  11 \\
\ce{CH3OH}(2$_{1,1}$--1$_{1,0}$)-E & 96755.507 & 38 & 0.262 & 28 & $0.20 \pm 0.04$ & $-4.70 \pm 0.13$ & $1.41 \pm 0.14$ & 56 &  11 \\
\ce{CH3OH}(2$_{1,1}$--1$_{1,0}$)-A & 97582.808 & 38 & 0.263 & 22 & $0.31 \pm 0.04$ & $-10.54 \pm 0.08$ & $1.26 \pm 0.08$ & 98 &  11 \\
\ce{CH3OH}(2$_{1,1}$--1$_{1,0}$)-A & 97582.808 & 38 & 0.263 & 22 & $0.28 \pm 0.06$ & $-4.94 \pm 0.16$ & $1.91 \pm 0.17$ & 57 &  11 \\
\ce{CH3OH}(11$_{-1,11}$--10$_{-2,9}$)-E & 104300.396 & 36 & 0.196 & 159 & $0.14 \pm 0.04$ & $-10.96 \pm 0.12$ & $0.96 \pm 0.12$ & 57 &  14 \\
\ce{CH3OH}(11$_{-1,11}$--10$_{-2,9}$)-E & 104300.396 & 36 & 0.196 & 159 & $0.27 \pm 0.05$ & $-5.03 \pm 0.11$ & $1.40 \pm 0.11$ & 77 &  14 \\
\ce{CH3OH}(0$_{0,0}$--1$_{-1,1}$)-E & 108893.929 & 34 & 1.471 & 13 & $0.95 \pm 0.09$ & $-10.72 \pm 0.07$ & $2.10 \pm 0.08$ & 180 &  15 \\
\ce{CH3OH}(0$_{0,0}$--1$_{-1,1}$)-E & 108893.929 & 34 & 1.471 & 13 & $0.26 \pm 0.07$ & $-4.23 \pm 0.20$ & $1.74 \pm 0.21$ & 59 &  15 \\
\\
\ce{CH3CN}($4_{2}$--$3_{2}$) & 73584.5431 & 50 & 1.639 & 37 & $0.12 \pm 0.04$ & $-10.87 \pm 0.31$ & $2.52 \pm 0.32$ & 20 &  7 \\
\ce{CH3CN}($4_{1}$--$3_{1}$) & 73588.7994 & 50 & 2.048 & 16 & $0.38 \pm 0.03$ & $-10.36 \pm 0.05$ & $1.60 \pm 0.06$ & 94 &  6 \\
\ce{CH3CN}($4_{0}$--$3_{0}$) & 73590.2183 & 50 & 2.185 & 9 & $0.34 \pm 0.03$ & $-10.24 \pm 0.05$ & $1.43 \pm 0.06$ & 94 &  7 \\
\ce{CH3CN}($5_{3}$--$4_{3}$) & 91971.1304 & 40 & 2.792 & 78 & $0.28 \pm 0.04$ & $-10.31 \pm 0.13$ & $2.24 \pm 0.15$ & 49 &  7 \\
\ce{CH3CN}($5_{3}$--$4_{3}$) & 91971.1304 & 40 & 2.792 & 78 & $0.10 \pm 0.03$ & $-3.71 \pm 0.26$ & $1.77 \pm 0.26$ & 23 &  7 \\
\ce{CH3CN}($5_{2}$--$4_{2}$) & 91979.9943 & 40 & 3.665 & 42 & $0.29 \pm 0.03$ & $-10.57 \pm 0.06$ & $1.61 \pm 0.06$ & 71 &  6 \\
\ce{CH3CN}($5_{2}$--$4_{2}$) & 91979.9943 & 40 & 3.665 & 42 & $0.11 \pm 0.03$ & $-4.45 \pm 0.15$ & $1.65 \pm 0.16$ & 28 &  6 \\
\ce{CH3CN}($5_{0}$--$4_{0}$) & 91987.0876 & 40 & 4.364 & 13 & $0.71 \pm 0.03$ & $-10.34 \pm 0.02$ & $1.52 \pm 0.02$ & 186 &  6 \\
\ce{CH3CN}($5_{0}$--$4_{0}$) & 91987.0876 & 40 & 4.364 & 13 & $0.75 \pm 0.03$ & $-4.52 \pm 0.02$ & $1.57 \pm 0.02$ & 190 &  6 \\
\ce{CH3CN}($6_{3}$--$5_{3}$) & 110364.3537 & 34 & 5.741 & 83 & $0.37 \pm 0.10$ & $-10.70 \pm 0.16$ & $1.82 \pm 0.17$ & 81 &  20 \\
\ce{CH3CN}($6_{3}$--$5_{3}$) & 110364.3537 & 34 & 5.741 & 83 & $0.17 \pm 0.07$ & $-4.45 \pm 0.21$ & $1.28 \pm 0.21$ & 54 &  20 \\
\ce{CH3CN}($6_{2}$--$5_{2}$) & 110374.989 & 34 & 6.805 & 47 & $0.29 \pm 0.06$ & $-10.35 \pm 0.10$ & $1.06 \pm 0.10$ & 109 &  22 \\
\ce{CH3CN}($6_{2}$--$5_{2}$) & 110374.989 & 34 & 6.805 & 47 & $0.11 \pm 0.09$ & $-4.93 \pm 0.45$ & $1.46 \pm 0.47$ & 29 &  22 \\
\ce{CH3CN}($6_{1}$--$5_{1}$) & 110381.372 & 34 & 7.444 & 26 & $0.82 \pm 0.08$ & $-10.39 \pm 0.05$ & $1.59 \pm 0.06$ & 206 &  17 \\
\ce{CH3CN}($6_{0}$--$5_{0}$) & 110383.4999 & 34 & 7.657 & 19 & $0.77 \pm 0.07$ & $-10.28 \pm 0.06$ & $1.49 \pm 0.06$ & 206 &  18 \\
\\
\ce{CH3CHO}($4_{1,4}$--$3_{1,3}$) & 74891.677 & 50 & 1.288 & 11 & $0.10 \pm 0.03$ & $-11.22 \pm 0.20$ & $1.94 \pm 0.20$ & 21 &  5 \\
\ce{CH3CHO}($4_{1,4}$--$3_{1,3}$) & 74924.1336 & 50 & 1.288 & 11 & $0.13 \pm 0.04$ & $-10.61 \pm 0.23$ & $2.40 \pm 0.23$ & 21 &  5 \\
\ce{CH3CHO}($4_{0,4}$--$3_{0,3}$) & 76866.4357 & 48 & 1.486 & 9 & $0.15 \pm 0.02$ & $-10.49 \pm 0.10$ & $1.41 \pm 0.10$ & 43 &  6 \\
\ce{CH3CHO}($4_{0,4}$--$3_{0,3}$) & 76878.9525 & 48 & 1.486 & 9 & $0.16 \pm 0.03$ & $-10.42 \pm 0.11$ & $1.69 \pm 0.11$ & 38 &  6 \\
\ce{CH3CHO}($5_{1,5}$--$4_{1,4}$) & 93580.9091 & 40 & 2.632 & 16 & $0.17 \pm 0.03$ & $-10.64 \pm 0.11$ & $1.69 \pm 0.11$ & 39 &  6 \\
\ce{CH3CHO}($5_{1,5}$--$4_{1,4}$) & 93595.2349 & 40 & 2.633 & 16 & $0.13 \pm 0.02$ & $-10.80 \pm 0.10$ & $1.35 \pm 0.10$ & 37 &  6 \\
\ce{CH3CHO}($5_{0,5}$--$4_{0,4}$) & 95947.4373 & 39 & 2.955 & 14 & $0.17 \pm 0.06$ & $-10.51 \pm 0.22$ & $1.87 \pm 0.22$ & 36 &  11 \\
\ce{CH3CHO}($5_{0,5}$--$4_{0,4}$) & 95963.4588 & 39 & 2.954 & 14 & $0.18 \pm 0.04$ & $-10.48 \pm 0.12$ & $1.33 \pm 0.12$ & 54 &  11 \\
\ce{CH3CHO}($5_{1,4}$--$4_{1,3}$) & 98863.3135 & 38 & 3.103 & 17 & $0.21 \pm 0.05$ & $-10.82 \pm 0.18$ & $1.90 \pm 0.18$ & 43 &  11 \\
\ce{CH3CHO}($5_{1,4}$--$4_{1,3}$) & 98900.9445 & 38 & 3.107 & 17 & $0.15 \pm 0.04$ & $-10.68 \pm 0.16$ & $1.54 \pm 0.16$ & 38 &  10 \\
\\
\ce{H2CCO}($4_{1,4}$--$3_{1,3}$) & 80076.652 & 46 & 0.504 & 23 & $0.14 \pm 0.03$ & $-10.24 \pm 0.13$ & $1.50 \pm 0.15$ & 37 &  6 \\
\ce{H2CCO}($4_{0,4}$--$3_{0,3}$) & 80832.117 & 46 & 0.553 & 10 & $0.06 \pm 0.02$ & $-10.68 \pm 0.13$ & $0.91 \pm 0.14$ & 25 &  5 \\
\ce{H2CCO}($4_{1,3}$--$3_{1,2}$) & 81586.23 & 45 & 0.533 & 23 & $0.13 \pm 0.02$ & $-10.31 \pm 0.09$ & $1.27 \pm 0.10$ & 41 &  5 \\
\ce{H2CCO}($5_{1,5}$--$4_{1,4}$) & 100094.514 & 37 & 1.030 & 27 & $0.16 \pm 0.03$ & $-10.26 \pm 0.09$ & $1.11 \pm 0.10$ & 58 &  9 \\
\ce{H2CCO}($5_{0,5}$--$4_{0,4}$) & 101036.63 & 37 & 1.104 & 15 & $0.06 \pm 0.03$ & $-9.80 \pm 0.29$ & $1.24 \pm 0.31$ & 19 &  10 \\
\ce{H2CCO}($5_{1,4}$--$4_{1,3}$) & 101981.429 & 36 & 1.089 & 28 & $0.19 \pm 0.07$ & $-10.56 \pm 0.26$ & $1.46 \pm 0.28$ & 53 &  15 \\
\\
\ce{t-HCOOH}($4_{1,4}$--$3_{1,3}$) & 86546.1891 & 43 & 0.635 & 14 & $0.13 \pm 0.03$ & $-9.96 \pm 0.18$ & $2.01 \pm 0.18$ & 26 &  6 \\
\ce{t-HCOOH}($4_{0,4}$--$3_{0,3}$) & 89579.1785 & 41 & 0.751 & 11 & $0.12 \pm 0.03$ & $-10.15 \pm 0.14$ & $1.34 \pm 0.14$ & 35 &  8 \\
\ce{t-HCOOH}($4_{2,2}$--$3_{2,1}$) & 90164.6296 & 41 & 0.575 & 24 & $0.06 \pm 0.02$ & $-10.34 \pm 0.23$ & $1.43 \pm 0.23$ & 17 &  6 \\
\ce{t-HCOOH}($4_{1,3}$--$3_{1,2}$) & 93098.3627 & 40 & 0.791 & 14 & $0.12 \pm 0.03$ & $-10.24 \pm 0.17$ & $1.58 \pm 0.17$ & 30 &  8 \\
\hline
\end{longtable}

\begin{longtable}
{lrrrrrrrrrr}
\caption{The detected lines used for optically thin/thick estimates of column densities and parameters of their HFS or Gaussian fits.}
\label{tab:detected_lines} \\
\hline
Line & $\nu$ & $\theta_b$ & $A$ & $E_l$ & $E_u$ & $W$ & $\tau$ & $V_{\rm LSR}$ & $\sigma$ & rms \\
& [MHz] & [$''$] & [$10^{-5}$ s$^{-1}$] & [K] & [K] & [K~km~s$^{-1}$] & & [km~s$^{-1}$] & [km~s$^{-1}$] & [mK] \\ \hline 
\ce{DCO^+}$(1-0)$ & 72039.3031 & 52 & 2.206 & 0 & 3 & $0.51 \pm 0.04$ & & $-10.44 \pm 0.08$ & $0.57 \pm 0.12$ & 17 \\ 
\ce{DCN}$(1-0)$ & 72414.6936 & 51 & 1.318 & 0 & 3 & $1.29 \pm 0.08$ & & $-10.54 \pm 0.07$ & $0.58 \pm 0.03$ & 23 \\ 
\ce{DNC}$(1-0)$ & 76305.7270 & 49 & 1.604 & 0 & 4 & & $4.58 \pm 2.03$ & $-10.06 \pm 0.07$ & $0.41 \pm 0.04$ & 16 \\ 
\ce{HC^{18}O^+}$(1-0)$ & 85162.2231 & 44 & 3.645 & 0 & 4 & $0.60 \pm 0.07$ & & $-10.42 \pm 0.13$ & $1.45 \pm 0.13$ & 29 \\ 
\ce{HCS^+}$(2-1)$ & 85347.8900 & 43 & 1.110 & 2 & 6 & $0.43 \pm 0.08$ & & $-10.49 \pm 0.20$ & $1.53 \pm 0.20$ & 33 \\ 
\ce{NH_2D}$(1_{1,1}-1_{0,1})$ & 85926.2630 & 43 & 2.056 \tablefootmark{a} & 18 & 21 & & $1.08 \pm 3.41$ & $-9.79 \pm 0.20$ & $0.24 \pm 0.07$ & 24 \\ 
\ce{HC^{15}N}$(1-0)$ & 86054.9664 & 43 & 2.203 & 0 & 4 & $1.21 \pm 0.05$ & & $-10.49 \pm 0.04$ & $1.59 \pm 0.04$ & 18 \\ 
\ce{H^{13}CN}$(1-0)$ & 86339.9214 & 43 & 2.225 & 0 & 4 & & $0.50 \pm 0.21$ & $-10.43 \pm 0.03$ & $0.65 \pm 0.01$ & 25 \\ 
\ce{H^{13}CO^+}$(1-0)$ & 86754.2884 & 43 & 3.853 & 0 & 4 & & $1.93 \pm 0.22$ & $-10.71 \pm 0.01$ & $0.52 \pm 0.01$ & 31 \\ 
\ce{HN^{13}C}$(1-0)$ & 87090.8500 & 43 & 1.867 & 0 & 4 & & $5.00 \pm 0.00$ & $-10.19 \pm 0.06$ & $0.43 \pm 0.01$ & 42 \\ 
\ce{H^{15}NC}$(1-0)$ & 88865.7150 & 42 & 1.983 & 0 & 4 & $0.51 \pm 0.03$ & & $-10.50 \pm 0.06$ & $1.51 \pm 0.06$ & 12 \\ 
\ce{t-HCOOH}$(4_{0,4}-3_{0,3})$ & 89579.1785 & 41 & 0.751 \tablefootmark{b} & 6 & 11 & $0.12 \pm 0.03$ & & $-10.40 \pm 0.25$ & $1.34 \pm 0.25$ & 14 \\ 
\ce{N^{15}NH^+}$(1-0)$ & 91205.9908 & 41 & 3.403 & 0 & 4 & $0.22 \pm 0.05$ & & $-9.99 \pm 0.39$ & $0.50 \pm 0.16$ & 18 \\ 
\ce{N_2H^+}$(1-0)$ & 93173.3977 & 40 & 3.628 & 0 & 4 & $24.80 \pm 0.25$ & & $-10.32 \pm 0.01$ & $0.52 \pm 0.01$ & 88 \\ 
\ce{CH_3OCHO}$(8_{3,6}-7_{3,5})$ & 98606.8740 & 38 & 1.198 \tablefootmark{b} & 23 & 32 & $0.26 \pm 0.03$ & & $-11.97 \pm 0.13$ & $1.08 \pm 0.13$ & 13 \\ 
\ce{CF^+}$(1-0)$ & 102587.4838 & 36 & 0.481 & 0 & 5 & $0.19 \pm 0.03$ & & $-11.29 \pm 0.15$ & $0.41 \pm 0.22$ & 18 \\ 
\ce{^{13}C^{18}O}$(1_{1/2}-0_{1/2})$ \tablefootmark{c} & 104711.3930 & 35 & 0.005 & 0 & 5 & $0.14 \pm 0.04$ & & $-10.68 \pm 0.23$ & $1.20 \pm 0.23$ & 22 \\ 
\ce{C^{18}O}$(1-0)$ & 109782.1734 & 34 & 0.006 & 0 & 5 & $16.33 \pm 0.15$ & & $-10.72 \pm 0.01$ & $1.45 \pm 0.01$ & 65 \\ 
\ce{^{13}CO}$(1-0)$ & 110201.3697 & 34 & 0.006 & 0 & 5 & $94.44 \pm 0.72$ & & $-10.01 \pm 0.01$ & $0.72 \pm 0.01$ & 298 \\ 
\ce{C^{17}O}$(1_{7/2}-0_{5/2})$ & 112358.9880 & 33 & 0.007 & 0 & 5 & & $0.81 \pm 0.28$ & $-10.72 \pm 0.03$ & $0.55 \pm 0.02$ & 42 \\   
\\
\ce{CH3OCH3}$ (3_{2,2}-3_{1,3})$ & 91476.6070 & 38 & 0.310 & 6 & 11 & $0.04$ & & & 2 & 6 \\
\ce{NH2CHO}$ (4_{1,3}-3_{1,2})$ & 87848.8735 & 42 & 4.299 & 7 & 14 & $0.06$ & & & 2 & 9 \\
\\
\ce{NH3}$ (1_{1,0,a}-1_{1,0,s})$ & 23694.4955 & & & & & & & $-10.37\pm0.08$ & $1.51\pm0.09$ & \\
\ce{NH3}$ (2_{2,0,a}-2_{2,0,s})$ & 23722.633335 & & & & & & & $-10.37\pm0.08$ & $1.51\pm0.09$ & \\
\ce{NH3}$ (3_{3,0,a}-3_{3,0,s})$ & 23870.1292 & & & & & & & $-10.37\pm0.08$ & $1.51\pm0.09$ & \\
\hline
\end{longtable}
\tablefoot{When $W$ is listed, the transition was modelled as optically thin; when $\tau$ is listed, the transition was modelled as optically thick. If a hyperfine fit was poor, $T_\text{ex}$ was replaced by $T_\text{ex}$(H$^{13}$CO$^+$). If the transition has no hyperfine structure, $T_\text{ex}$ was also assumed to be $T_\text{ex}$(H$^{13}$CO$^+$). We assumed a source size of 80'' except for \ce{H^13CN}, \ce{HC^15N}, \ce{HN^13C} and \ce{H^15NC} where we assumed a source size of 107$''$. 
\tablefoottext{a}{Retrieved from \cite{Petrashkevich_2024} 
(not calculated because \ce{NH2D} is not a linear molecule).}
\tablefoottext{b}{Retrieved from CDMS/JPL \citep{Pickett1998,Muller2001,Endres2016}, because the molecules are not linear. }
\tablefoottext{c}{The frequency difference between the hyperfine components of \ce{^13C^18O} is $\sim$ 46.4 kHz, which is smaller than the channel size, so \ce{^13C^18O} is modelled as a single line. }}

\end{center}
\end{landscape}

\end{document}